\documentclass[a4paper,aps,twocolumn,preprintnumbers]{revtex4}

\usepackage{graphicx}  % Include figure files
\usepackage{subfigure}
\usepackage{multirow}
\usepackage[colorlinks,linkcolor=blue,citecolor=blue]{hyperref}
\usepackage{siunitx}
\usepackage{float}

\usepackage{fancyhdr}
\usepackage{longtable}
\usepackage{parskip}
\usepackage[T1]{fontenc}
\usepackage{dcolumn}   % Align table columns on decimal point

\usepackage{bm}        % bold math
\usepackage{amsfonts}  % Common math fonts
\usepackage{amsmath}   % Common math functions
\usepackage{amssymb}   % Common math symbols
\usepackage{lmodern}
\usepackage[inkscapearea=page]{svg}

\newcommand{\pwisein}{\left\{ \begin{array}{ll}}
\newcommand{\pwiseout}{\end{array}\right.}

\begin{document}

\title{Demonstration of a Multiplexing Trapped Ion Quantum Processing Unit}

\author{F. Anmasser$^{1,2}$, M. Abu Zahra$^{3,4}$, K. Schüppert$^{2}$, M. Pototschnig$^{2}$, J. Wahl$^{1}$, M. Dietl$^{1,2}$, M. Pfeifer$^{1,2}$,  Y. Colombe$^{2}$, J. Repp$^{3}$, M. Brandl$^{3}$, P. Schindler$^{1}$ and C. Rössler$^{2}$}

\affiliation {
\it 
$^{1}$ Institut für Experimentalphysik, Universität Innsbruck, Technikerstraße 25, A-6020, Innsbruck, Austria \\
$^{2}$ Infineon Technologies Austria AG, Siemensstraße 2, A-9500, Villach, Austria \\
$^{3}$ Chair of Circuit Design, Technical University of Munich, Arcisstraße 21, 80333, Munich, Germany \\
$^{4}$ Infineon Technologies AG, Am Campeon 1-15, 85579, Neubiberg, Germany
}

\date{\today}

% new abstract (shorter, more specific)
\begin{abstract}
A fault-tolerant quantum computer is expected to require thousands of qubits. Trapped ion architectures provide a modular approach where the quantum register is divided into multiple subregisters connected by physically moving the corresponding ions. Transporting ions at scale comes with several challenges such as the need to connect thousands of control lines to an ion trap chip. Multiplexing the required control voltages from few input signals to multiple electrodes offers a solution to this wiring challenge. Here we demonstrate a quantum processing unit that combines a surface ion trap with a time multiplexer via a sample-and-hold technique that initially charges electrodes to fixed voltages and disconnects them during qubit operations. We characterize the unit's performance by measuring motional heating rates below one phonon per second in both open and closed switch configurations. We further characterize the sample and hold process and find that sampling intervals below 50 ms are sufficient to keep expected gate errors from decaying charges during the hold phase below $10^{-4}$. Our results indicate that the multiplexing scheme is compatible with high-fidelity operations. 
%A fault-tolerant quantum computer is expected to require thousands of qubits, but scaling trapped-ion systems is severely limited by the need for one control line per electrode. On-chip electronic multiplexing offers a solution by substantially reducing wiring requirements. Here we demonstrate a quantum processing unit combining a surface ion trap with an electronic multiplexer operating in a time-multiplexed sample-and-hold mode, where electrodes are charged to target voltages and left floating during qubit operations. We characterize the voltage decay on floating electrodes and calculate that sampling intervals below 50 ms are sufficient to keep expected gate errors below $10^{-4}$. We further measure motional heating rates below one phonon per second in both open and closed switch configurations, demonstrating that multiplexed control introduces no substantial electrode noise. These results show that on-chip multiplexing is compatible with high-fidelity trapped-ion operation, representing a key step toward wiring-scalable quantum computers.
\end{abstract}

% old abstract
% \begin{abstract}
% A fault-tolerant quantum computer is expected to require thousands of qubits to solve problems that are intractable for classical computers. Trapped-ion technology, which encodes qubits in charged particles confined above planar electrodes, is a leading platform for such scalable quantum computing systems. However, each ion requires multiple independently controlled electrodes, and the number of required control lines scales directly with the qubit count, leading to a severe wiring bottleneck. On-chip electronic multiplexing offers a potential solution to this challenge by substantially relaxing wiring requirements. Here we demonstrate a quantum processing unit consisting of a surface ion trap and an electronic multiplexer capable of time-multiplexed sample-and-hold operation, in which electrodes are first charged to their target voltages and subsequently left floating while quantum gates are performed. We determine that sampling intervals below approximately \SI{50}{ms} are sufficient to maintain stable trapping potentials and keep expected gate errors below $10^{-4}$. We further characterize the QPU's switching behavior by examining charge injection and voltage decay on floating electrodes with respect to ion trapping performance. Specifically, we measure motional heating rates below one phonon per second for both open and closed switch configurations, demonstrating that multiplexed control can be implemented without adding substantial noise to the electrodes.
% \end{abstract}

\maketitle 

\tableofcontents

\section{Introduction}
\label{sec:introduction}

%Trapped ions platform
Trapped ions are considered as a promising platform for quantum simulation and quantum computation. Single qubit operations with an error rate in the order of $10^{-7}$ and coherence times of several thousand seconds have been achieved, along with two-qubit gate operations exhibiting an error rate below $10^{-4}$ \cite{smith2024singlequbitgateserrors107, Wang_2021, löschnauer2024scalablehighfidelityallelectroniccontrol, Hughes2025TrappedionTG}. Current state-of-the-art quantum processors based on trapped ions are capable of performing arbitrary quantum circuits with up to around 100 qubits \cite{PhysRevX.13.041052, Chen2023BenchmarkingAT, Pogorelov2021CompactIQ, alam2025fermionicdynamicstrappedionquantum, ransford2025helios98qubittrappedionquantum}, and may be scaled up to a qubit count in the several hundreds \cite{Preskill2018QuantumCI, Bharti2021NoisyIQ}. The number of qubits required for a universal fault-tolerant quantum computer is still an active area of research. Although arlier estimates suggested that certain problems would require upwards of a million qubits \cite{Gidney_2021}, recent advances in error correction protocols indicate that a few tens of thousands of qubits may be sufficient \cite{tripier2026faulttolerantquantumcomputingtrapped, Cain2026ShorsAI}.
% Estimates for certain problems suggest that above a million qubits are needed \cite{Gidney_2021}.

%Advantages of surface ion traps
Reaching large qubit numbers in a single device can be facilitated by a two-dimensional trapping geometry, making surface ion traps the preferred architecture \cite{Kielpinski2002ArchitectureFA, Seidelin_2006}. Surface ion traps are compatible with established semiconductor microfabrication processes, enabling precise and reproducible fabrication. Such compatibility provides significant potential for scaling chip traps in terms of qubit count \cite{Romaszko2019EngineeringOM, Auchter2022IndustriallyMI, Badawi2025ChipletTF}.

\begin{figure*}
\includesvg[scale=0.29]{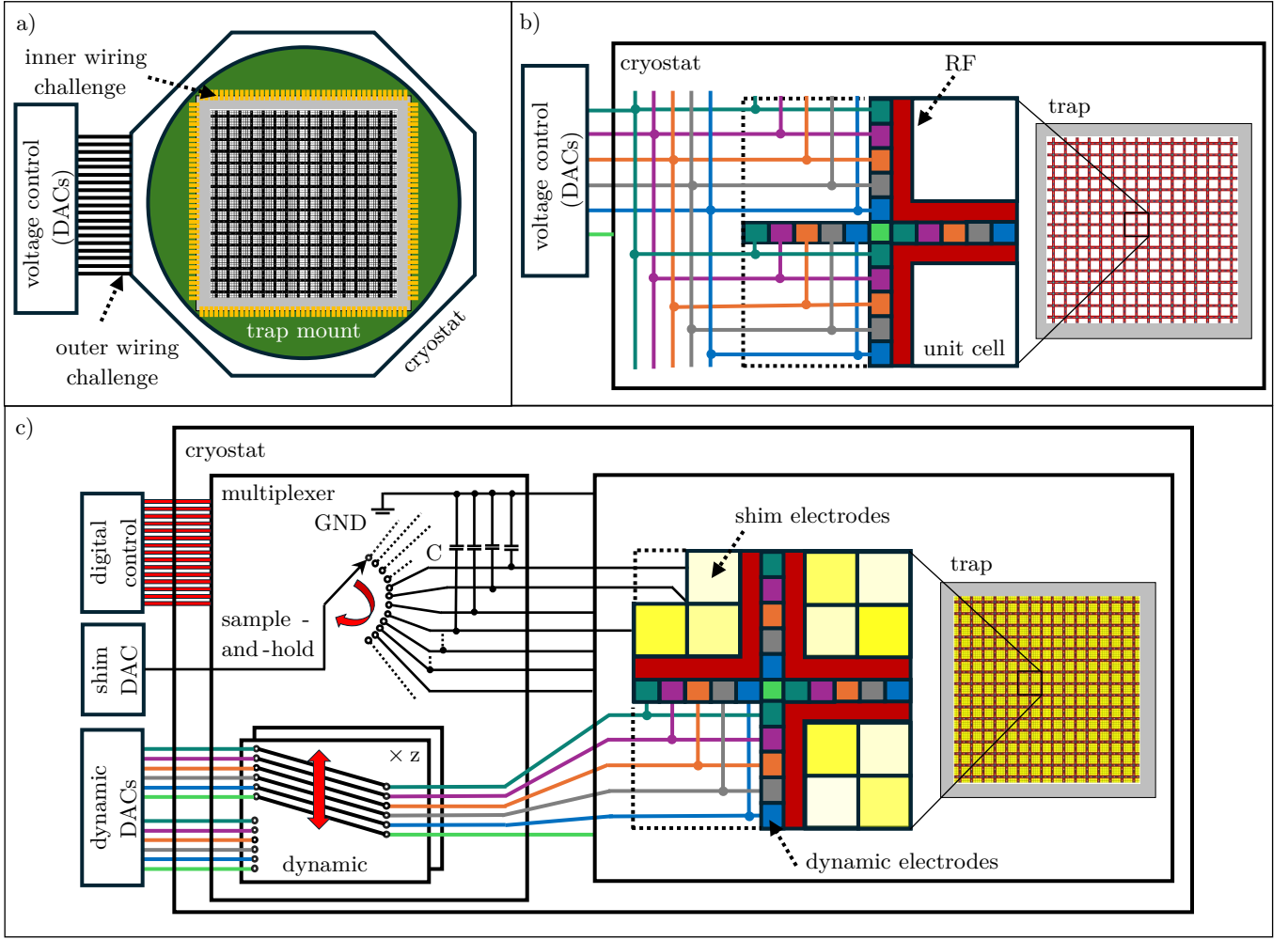}
\caption{\label{fig:architect} Illustration of the architectural complexities of trapped-ion quantum computing systems. a) DACs are situated outside the cryostat and every electrode is connected to a single DAC. b) Electrode co-wiring: electrodes with same functionality in different unit cells are connected to the same DAC outside the cryostat. c) Integrated switching electronics: a multiplexing unit is placed between the DACs and the chip trap within the cryostat. The DACs and electrodes are categorized into two types: dynamic and compensation electrodes. }
\end{figure*}

%Architectures and ion crystal reconfiguration
Ions confined within a potential well naturally arrange into ordered structures known as ion crystals \cite{vanMourik2021rfinducedHD, Rohde2000SympatheticGC, Mao2021ExperimentalRO}, but a single crystal is not scalable to the large numbers of qubits required for practical quantum computing applications. To utilize the properties of ion crystals for quantum computing, various architectures for chip traps \cite{Kielpinski2002, Pino2020DemonstrationOT, Mordini2024MultizoneTQ, RyanAnderson2021RealizationOR, PhysRevX.13.041052, Moses2020DemonstrationOT, Paetznick2024DemonstrationOL, Valentini2024, Monroe_2014}, most prominently the Quantum-Charged-Coupled-Device (QCCD), have been explored theoretically and experimentally. In the QCCD architecture, each trapping zone hosts a linear ion crystal, functioning as a quantum subregister. Gate operations within a single crystal are straightforward implementable because all ions share the same motional modes. However, interactions between qubits located in different subregisters cannot be performed directly. Instead, ion registers must be reconfigured by physically moving the ions. The reconfiguration of ion crystals requires typically ion transport, swapping of neighboring ions, and splitting and merging of ion crystals \cite{Kaushal2019ShuttlingbasedTQ, Mourik2020CoherentRO}.

%Scaling of wires
In surface ion traps, a combination of voltages in the radio frequency (RF) regime and DC voltages \footnote{Although the term 'DC' typically implies a constant direct current, in the context of surface ion traps, it is widely used to refer to the time-dependent, quasi-static voltages applied to the electrodes for axial confinement. However, it is worth noting that this terminology is slightly misleading, as the voltages are not truly constant and do not involve a direct current, but rather a time-varying electric potential. Despite this technical inaccuracy, the term 'DC' has become an established convention in the field, and we adopt it here for consistency and clarity.} are used to confine the ions above the surface of the trap. Typically, a single RF voltage source and two RF electrodes are providing confinement for all ions of the trap in two spatial directions. Confinement in the third direction is realized by the DC electrodes. Precise control of a potential well requires at least eight independently driven DC electrodes, corresponding to the eight independent spherical harmonic modes up to second order \cite{Mourik2020CoherentRO}. For practical operation in a realistic device, including the splitting and merging of ion crystals, approximately ten electrodes per potential well provide a reasonable estimate \cite{Kaushal2019ShuttlingbasedTQ, Moses2020DemonstrationOT, PhysRevX.13.041052}. Each potential well can host up to approximately ten ions forming a subregister \cite{Monz_2011}. Since each subregister requires to be controlled individually, the number of connections scales linearly with the number of subregisters. With ten electrodes and ten qubits per subregister, the number of wires needed is in the same order of magnitude as the qubit count. This reasoning implies that a processor with a million qubits would require about a million separate control lines.

%Cryostat and wiring challenges
These connection requirements become increasingly critical in the context of the system infrastructure. Chip traps are operated within ultra-high vacuum chambers to eliminate collisions between trapped ions and environmental gas molecules, a necessary condition for enhancing the lifetime of trapped ions \cite{Wineland2003}. Cryogenic temperatures are essential for reaching the highest attainable vacuum levels \cite{Schwarz2012CryogenicLP} and further reduce gate error rates by minimizing surface noise of the chip trap \cite{Bruzewicz2014MeasurementOI, Labaziewicz2008TemperatureDO}. The voltage for each electrode is supplied through an electrical connection, which is routed from the trap chip to the outside of the chamber, where digital-to-analog converters (DACs) provide the necessary voltages. As the number of qubits grows, two notable wiring challenges emerge at two distinct interfaces, as shown in Figure \ref{fig:architect}\textcolor{blue}{a}. Firstly, the interface between the external environment and the interior of the cryostat becomes a critical point of consideration, since maintaining thermal isolation becomes unfeasible  when roughly a million lines must penetrate the cryostat. Secondly, the interface between the cryostat interior and the trap chip forms a significant interconnect constraint. This arises from the need to minimize chip trap areas in order to minimize ion‑transport distances. Individual DC electrodes typically occupy areas of $100 \times 100 ~\mu \mathrm{m}^2$ or less, and the corresponding density of electrical interconnects not only strains current technologies \cite{Chen2019SystemOI, Vivet2021IntActA9}, but also leaves insufficient space for other crucial components such as RF routing, and optical waveguides or microwave routing structures for qubit control \cite{Mehta2020IntegratedOptical, Niffenegger2020IntegratedMultiwavelength, Sutherland2019}.

%Possible solutions
Several strategies to reduce the number of required connections have been proposed, including i) co-wiring of control electrodes \cite{Holz20202DLT, Alonso_2013, Seidler2022, PhysRevX.13.041052}, ii) integration of DACs into the cryogenic system \cite{Stuart2019, Meyer2023A1B, Sieberer2024ACD, Park2024AC6, Meyer2025CryogenicEO}, and iii) integration of multiplexers consisting of analog switches into the cryogenic system \cite{Malinowski2023, Delaney2024}. We analyze these approaches with respect to the cryostat’s thermal shielding, the heat load, footprint and needed data bandwidths of integrated electronics, and the ability to perform arbitrary ion‑crystal reconfigurations.

\subsection{Co-wiring of Electrodes}

The co-wiring approach physically connects multiple electrodes to the same DAC, reducing the number of required connections. Electrodes in different regions of the chip, but with the same functionality, are connected to the same DAC \cite{Holz20202DLT, Alonso_2013, Seidler2022, PhysRevX.13.041052}, as indicated in Figure \ref{fig:architect}\textcolor{blue}{b}. In such an architecture, it is required to base the trap layout on a unit cell \cite{Malinowski2023, Delaney2024}. Connecting electrodes of different unit cells to a common control wire reduces the ratio of wires to ions. This architecture enables the simultaneous transport of multiple ion crystals across different cells. However, due to the static routing, the crystals are locked into their respective positions, following a fixed sequence that cannot be dynamically rearranged. This rigidity prevents the ability of arbitrary ion crystal reconfigurations. Additionally, a globally controlled architecture like co-wiring, cannot compensate for local perturbations in the confining potentials. Such perturbations are unavoidable in real devices and arise from variations in control voltages due to finite line resistances as well as from local stray electric fields \cite{Lee2023PhotoinducedCD}.

\subsection{Cryogenic Integrated DACs}

Another way to address the wiring challenge is to integrate the voltage sources directly with the trap chip.  Integrating DACs inside the cryostat reduces the number of wires routed into the cryostat \cite{Stuart2019, Meyer2023A1B, Sieberer2024ACD, Park2024AC6, Meyer2025CryogenicEO}. Commercial DACs have been successfully incorporated into cryogenic systems \cite{Guise2014} and custom DACs have been monolithically integrated into an ion trap \cite{Stuart2019}. State-of-the-art DACs, operting at cryogenic temperatures dissipate about 1 - \SI{30}{mW} per channel \cite{Sieberer2024ACD}, which is detrimental to maintaining cryogenic conditions with modern cryostats providing only \SI{1.5}{W} or less of cooling power \cite{Spivey2021HighStabilityCS, Micke2019ClosedcycleL4, Hartsell2025DesignAC}, limiting the DAC count to around 150. In addition to their thermal load, the per-DAC footprints of about $\SI{0.035}{mm^2}$ \cite{Stuart2019} are oversized relative to DC electrodes, which are typically around $\SI{0.01}{mm^2}$ or smaller. Moreover, supporting individually driven DACs is expected to require data bandwidths approaching \SI{100}{Mbit/s} per qubit to deliver the waveforms necessary for ion‑crystal reconfiguration \cite{Malinowski2023}. These constraints restrict the scaling of architectures based on integrated‑DACs.

\subsection{Cryogenic Integrated Multiplexers}
\label{subsec:WISE}

An analog multiplexer is a switching device that takes a single analog input signal and routes it to one of several output lines or vice versa. Integrating analog multiplexers within the cryostat, rather than DACs, offers a potential route toward scaling up qubit numbers in quantum processing units. This approach offers the benefits of small footprints (< $\SI{0.01}{mm^2}$ per switch \cite{Stuart2019, Zahra2024EvaluationOL}), low power dissipation ($\SI{39}{\mu W}$ per switch \cite{Zahra2024EvaluationOL}), and low data bandwiths (\SI{50}{kbit/s} per qubit \cite{Malinowski2023}). The architectures introduced in References \cite{Malinowski2023, Delaney2024} make use of integrated switching electronics and distinguish between two types of electrodes: dynamic electrodes and compensation electrodes, see Figure \ref{fig:architect}\textcolor{blue}{c}. 

Dynamic electrodes are co-wired and deliver waveforms that enable ion crystal reconfiguration across multiple cells on the chip simultaneously. To achieve arbitrary reconfiguration of ion crystal chains, both the ability to maintain the position of individual crystals and the ability to swap the positions of adjacent crystals are required, as these two operations enable the rearrangement of ions into any desired configuration \cite{cormen2009introduction}. Therefore, the dynamic electrodes are connected through switches to two separate sets of DACs, as shown in the bottom of the multiplexer panel in Figure \ref{fig:architect}\textcolor{blue}{c}. The first set delivers waveforms to perform ion crystal reconfiguration, while the second one delivers constant voltages, maintaining the crystals in their positions. This configuration is repeated in each unit cell, where the state of the switches determines whether the ions within that cell undergo reconfiguration or remain stationary in their current positions. While dynamic electrodes enable global control and reconfiguration of ion crystals, local electric field perturbations still need to be compensated, which is achieved through the use of compensation electrodes.
% The states of the switches determines whether a reconfiguration operation is performed within a specific zone or not and controlling the switches for the dynamic DACs enables arbitrary reconfiguration of ion crystals. 

% compensation electrodes and sample and hold
These compensation electrodes are biased to specific potentials using a multiplexing approach. To this end, analog switches implement a time-multiplexing scheme, as illustrated in Figure \ref{fig:architect}\textcolor{blue}{c} (multiplexer panel, top). Electrodes are modeled as one plate of a capacitor, with the common ground as the opposing plate. When the voltage source is disconnected, the electrode voltage decays with a time constant inversely proportional to its capacitance to ground. This decay can be exploited by connecting multiple electrodes to a single DAC via a multiplexer. Multiple electrodes can be biased to different potentials using a single DAC is achieved through a 'sample-and-hold' \cite{Stuart2019, Malinowski2023, Delaney2024} procedure. This involves applying a voltage to an electrode via the DAC and opening the switch to isolate the electrode. The next electrode is then connected to the DAC and biased to a certain potential. That sequence of connecting, charging, and disconnecting is repeated for each electrode. To ensure low enough decay rates, sufficiently large capacitances must be incorporated into the trap chip or multiplexer. 

In summary, the three strategies presented above exhibit distinct trade-offs with regard to mitigating the wiring bottlenecks in large-scale trapped ion processors. Co-wiring minimizes connections and preserves the cryostat’s thermal isolation but sacrifices flexibility, as global control prevents compensation of local field perturbations and precludes arbitrary ion‑crystal reconfigurations. Hence, the co-wiring architecture alone is not scalable. Integrating DACs inside the cryostat reduces the number of cryostat feedthroughs, yet their substantial heat dissipation, large device footprints, and demanding data‑bandwidth requirements render this approach difficult to scale. In contrast, cryostat‑integrated analog multiplexers offer substantially lower heat loads, require lower bandwiths, and smaller footprints while retaining the ability to perform ion‑crystal reconfigurations through a combination of dynamic and sample‑and‑hold electrodes. This architecture seems to provide the most balanced route toward scalability.

Designing and fabricating multiplexers compatible with cryogenic temperatures is crucial for architectures using integrated switches. Suitable analog switches integrated into the chip trap \cite{Stuart2019} or on a separate printed circuit board inside the cryogenic chamber \cite{Home2016} have been successfully combined with chip traps before. However, a multiplexer for large-scale trapped ion processors constitutes not merely a collection of analog switches but a coordinated architecture that incorporates control logic, addressing circuitry and integrated capacitances. Such a multiplexing system to address the wiring challenge has not been demonstrated so far to our knowledge. 

In our study we introduce a modular unit consisting of a surface ion trap and a multiplexer, which we denote quantum processing unit (QPU), outlining its design, fabrication, and characterization. The QPU presented in this work  consists of an ion trap and a multiplexer, both operated at cryogenic temperatures. The details of the fabricated hardware are discussed in Section \ref{sec: qpu}. Section \ref{sec:results} presents the results of the QPU characterization with trapped ions, highlighting its performance and functionality with regard to heating rates at different metal temperatures and the sample-and-hold technique. Finally, Section \ref{sec:outlook} provides a forward-looking analysis, outlining strategies and estimations for scaling multiplex-based QPU architectures.

\section{Hardware}
\label{sec: qpu}

% introduction to this chapter
This section presents the design, fabrication, and assembly of the quantum processing unit (QPU) used in this study. The QPU consists of two main components: the multiplexer and the surface ion trap. The multiplexer is a critical component that enables the control of the ion trap's electrodes, and its specifications and performance are summarized in References \cite{Zahra2024EvaluationOL, Zahra2025}. The surface ion trap is another key component, and its design and fabrication are discussed in the following subsection. Finally, the assembly of the two chips on a carrier PCB and the experimental cryogenic setup are presented.

\subsection{Multiplexer}
\label{subsec:multiplexer}

\begin{figure*}
\centering
\includesvg[scale=0.29]{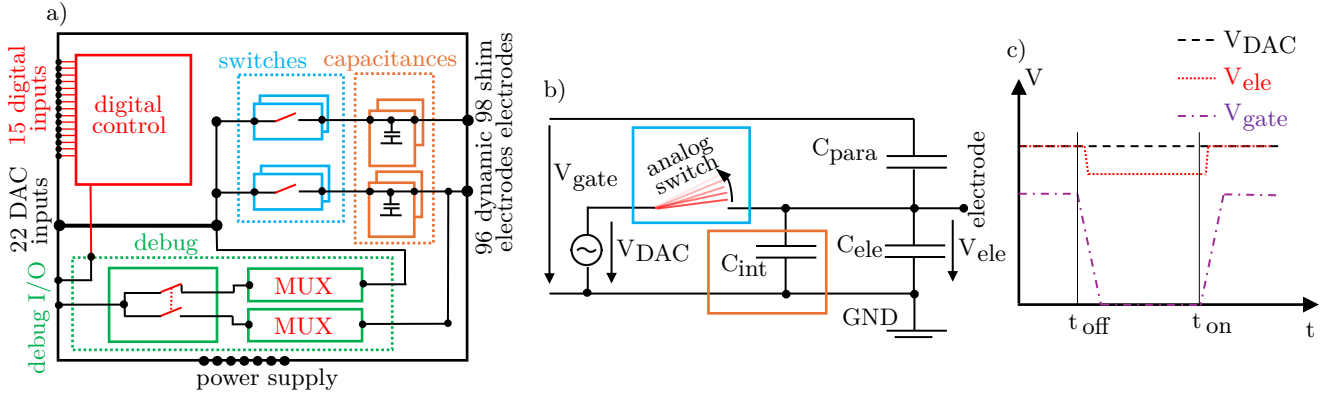}
\caption{\label{fig:charge_inject} 
a) High level block diagram of the multiplexer. Inputs of the multiplexer are the DACs, digital control and power supplies. The configuration of open and closed switches is controlled by the digital control. Capacities are integrated into the multiplexer. There are 96 outputs for dynamic electrodes and 98 outputs for shim electrodes. A dedicated function for debugging is able to multiplex any input voltage back to the outside of the cryostat. b) Simplified schematic of the switch and electrode circuit, highlighting the key components involved in the charge injection mechanism. c) Sketched diagram illustrating the timing of charge injection and its effect on the electrodes potential. 
}
\end{figure*}

% generalities
The basis for this work is an application-specific integrated circuit (ASIC) multiplexer, developed and fabricated by Infineon Technologies AG \cite{Zahra2024EvaluationOL, Zahra2025}. A digitally controlled switching matrix enables the distribution of 22 DAC inputs across 199 DC electrodes. The chip features integrated capacitances. Additionally, a debug function for monitoring its state during operation is incorporated to facilitate diagnostic testing. Figure \ref{fig:charge_inject}\textcolor{blue}{a} shows a high level block diagram of its functionalities. A detailed characterization of the ASIC at temperatures ranging from \SI{300}{K} to \SI{4}{K} has been reported in previous studies \cite{Zahra2024EvaluationOL, Zahra2025}. The chip is based on Infineon's \SI{130}{nm} bipolar transistor technology and has dimensions of $\SI{13}{\mathrm{mm}} \times \SI{6}{\mathrm{mm}}$. The multiplexer’s voltage range, both input and output, is limited to $\SI{\pm 10}{\mathrm{V}}$. The multiplexer features integrated capacitances, a crucial aspect for sample-and-hold applications, with values of \SI{15}{pF} for dynamic electrodes and \SI{50}{pF} for shim electrodes.

% sample and hold and charge injection
A crucial aspect of analog switch performance is the phenomenon of charge injection, which occurs during their switching dynamics \cite{chargeinjection}. This charge injection is caused by parasitic charges, which are injected onto the electrode, thereby perturbing its voltage and leading to a sudden change in the confinement of the ion. In Figure \ref{fig:charge_inject}\textcolor{blue}{b}, an equivalent circuit diagram is shown. Here, the output voltage of the DAC, $V_{\mathrm{DAC}}$, is connected via an analog switch to the electrode, which forms a capacitor, $C_{\mathrm{ele}}$, to ground. In parallel to the electrode capacitor, there is the integrated capacitance of the multiplexer $C_{\mathrm{int}}$. The switch is turned on or off by the gate voltage, $V_{\mathrm{gate}}$. The physical structure of the transistors, with multiple layers of conductive and insulating materials, creates a parasitic capacitance $C_{\mathrm{para}}$ (around \SI{1}{pF} for the devices used in this work). Initially, when the switch is closed, the electrode is at a voltage of $V_\mathrm{ele} = V_{\mathrm{DAC}}$, with $V_\mathrm{gate} > 0$. As a result, the charges stored in the two capacitors can be calculated as,

\begin{equation}
\begin{split}
    Q_{\mathrm{para}} &= (V_{\mathrm{DAC}} - V_{\mathrm{gate}})\,C_{\mathrm{para}}, \\ Q_{\mathrm{ele}} &= V_{\mathrm{DAC}}\, (C_{\mathrm{ele}} + C_{\mathrm{int}}).
\end{split} 
\end{equation}
After the switch is opened, the charge stored in each capacitor is rebalanced among them, leading to new charge values,

\begin{equation}
\begin{split}
    Q_{\mathrm{para}}^{\prime} &= V_\mathrm{ele}^{\prime}\,C_{\mathrm{para}}, \\
    Q_{\mathrm{ele}}^{\prime} &= V_\mathrm{ele}^{\prime} \,(C_{\mathrm{ele}} + C_{\mathrm{int}}).
\end{split} 
\end{equation}
Given that the total charge remains conserved during the transition from a closed to an open switch state, $Q_{\mathrm{para}} + Q_{\mathrm{ele}} = Q_{\mathrm{para}}^{\prime} + Q_{\mathrm{ele}}^{\prime}$, we can derive an expression for the electrode's voltage after the switch is open,

\begin{equation}
    V_{\mathrm{ele}}^{\prime} = \frac{V_{\mathrm{DAC}} \,(C_{\mathrm{ele}} + C_{\mathrm{int}}) + (V_{\mathrm{DAC}} - V_{\mathrm{gate}}) \,C_{\mathrm{para}}}{C_{\mathrm{para}} + C_{\mathrm{ele}} + C_{\mathrm{int}}}.
    \label{equ:charge_injection}
\end{equation}
The transition from the closed switch state to the open state with the change in the electrode potential is sketched in Figure \ref{fig:charge_inject}\textcolor{blue}{c}. The magnitude of the change in potential is determined by the applied voltage. For large $C_\mathrm{int}$ or $C_\mathrm{ele}$, the effect of charge injection onto the electrode can be suppressed. Electrical tests of the multiplexer revealed potential drops of about \SI{10}{\%} at the maximum nominal voltage of \SI{10}{V} \cite{Zahra2025}.

In addition to toggling switches, a second key factor for scalable multiplexing architectures is the voltage decay rate of electrodes with open switches. After transitioning the switch from closed to open, measurements at \SI{4}{K} showed a voltage decay rate of \SI{0.47}{mV/s} for \SI{5}{V} nominal voltage and \SI{50}{pF} \cite{Zahra2025}. In the present study, we measure the effect of charge injection and the voltage decay rate for floating electrodes directly on trapped ions. 

\subsection{Surface ion trap}

\begin{figure}
\centering
\includesvg[scale=0.26]{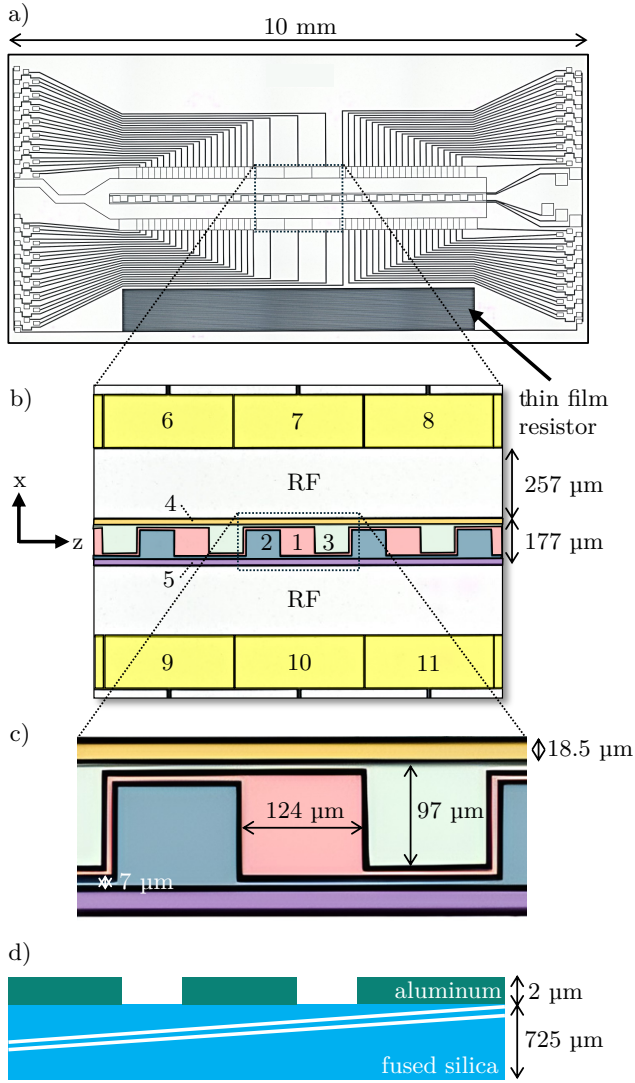}
\caption{\label{medusa_layout}Trap design. a) Stitched microscope image of the surface ion trap on a fused silica substrate, with chip dimensions of \SI{10}{mm} $\times$ \SI{5}{mm}. The bottom gray rectangle is a thin film resistor connected to the bond pads. Bond pads for connecting the electrodes and multiplexer via wire bonds are arranged in three columns on the left and right sides of the chip. b) Five DC electrodes, nested in between the RF rails. Three of these electrodes have a 'zig-zag' shape to connect to the bond pads. Compensation electrodes, shown in yellow, are located outside the RF rails. c) Geometry of the zig-zag electrodes. d) Cross-section sketch of the ion trap, featuring a single electrode metal layer deposited on a fused silica substrate.}
\end{figure}

\subsubsection{Trap design}

% height motivation
In designing our surface ion trap, a primary considerations was the ion-surface distance $d$. For ion-surface distances below $\SI{100}{\mu \mathrm{m}}$, laser stray fields can generate uncontrolled electric fields, compromising trap stability. To facilitate reliable ion trapping with the multiplexer, we selected $d=\SI{170}{\mu \mathrm{m}}$.

% DC limits
In addition to the ion-surface distance, the maximum DC voltages required to confine ions at frequencies in the MHz range were a crucial consideration. The maximum DC voltages necessary for ion confinement scale directly with the ion-surface distance. Hence, a closer proximity between the ion and the surface is preferable, given the $\SI{\pm 10}{V}$ output voltage constraint of the multiplexer.

The presented design (Figure \ref{medusa_layout}\textcolor{blue}{a-c}) allows for $d=\SI{170}{\mu \mathrm{m}}$ while operating within $\SI{\pm 10}{V}$, using a single metal layer. A distinctive feature of the design are the DC electrodes in a 'zigzag' shape, nested in between the RF rails to minimize the distance between the surface and the ions. The zigzag shape of electrodes 1, 2 and 3 enables routing to the bond pads on the right-hand side of the chip, using a single metal layer. The interconnecting traces have a width of $\SI{7}{\mu m}$ and are separated by $\SI{3}{\mu m}$ from adjacent electrodes. Two DC electrodes (numbered 4 and 5 in Figure \ref{medusa_layout}\textcolor{blue}{b}) with a width of $\SI{18.5}{\mu m}$ are placed in between the RF rails. Ion transport voltage profiles and voltage shim sets for the inner DC electrodes (1-5), confining the ion at any point along the trap path, are given in Appendix \ref{app:shim}. With only five electrodes available, there are insufficient degrees of freedom to compensate for local stray fields when multiple ions are distributed across different regions of the chip. Two rows of compensation electrodes at the outside of the RF rails compensate for local stray fields along the RF null. However, in order to reduce experimental complexity, only the compensation electrodes 6-11 were connected to DACs for this study. All other compensation electrodes were connected to ground potential, which is the unstructured top metal area on the chip. Two RF rails, each $\SI{257}{\mu m}$ wide and separated by $\SI{177}{\mu m}$, confine the ions along a \SI{6}{mm} path in the axial (z) direction. 

The trap design incorporates structures, which are able to heat and measure the temperature of the electrode metal layer. These structures are meandering, continuous paths, forming a thin film resistor \cite{Dietl2025}. The thin film resistor is filling out the majority of the free space available (see Figure \ref{medusa_layout}\textcolor{blue}{a} or Figure \ref{qpu} in the North (N) and South (S) of the trap). The overall path of the thin film resistor is \SI{75}{cm} long and has a thickness of $\SI{2}{\mu m}$, resulting in a measured resistance of $\SI{3.2}{\mathrm{k} \Omega}$ at room temperature. The thin film resistor is able to heat up the trap by applying a voltage or forcing a current on both ends. The temperature-dependent resistance of the metal allows the thin film resistor to also function as a temperature sensor.

Variations of the design, which differed only in minor aspects, were utilized for this work. The trap in Figure \ref{medusa_layout}\textcolor{blue}{a} features one thin film resistor and 96 compensation electrodes. The traps shown in Figure \ref{qpu} accommodate two thin film resistors, but have six compensation electrodes, reducing experimental complexity. A third design with an ion height of $\SI{80}{\mu m}$ is presented in Section \ref{sec:outlook} and is intended for future experiments in conjunction with the multiplexer.

\subsubsection{Fabrication}

% stack
The ion traps were fabricated at Infineon Technologies' industrial facilities in Villach, Austria. Chip traps fabricated on silicon substrates exhibit a factor of 3.3 to 5.6 higher RF power dissipation, compared to fused silica \cite{Dietl2025}. Hence, fused silica was chosen as substrate material, which is present in the form of disks, measuring \SI{200}{mm} in diameter and having a thickness of $\SI{725}{\mu \mathrm{m}}$. Each wafer provides space for approximately 450 ion traps. Ion traps with several metal routing layers have been fabricated at Infineon in Villach \cite{Auchter2022IndustriallyMI, Dietl2025}, but we chose a simplified fabrication process, facilitating more frequent design iteration cycles. Consequently, the trap consists of a single aluminum layer on top of a fused silica substrate. A cross-sectional view is shown in Figure \ref{medusa_layout}\textcolor{blue}{d}.

% sputtering
Via physical vapor deposition (PVD) \cite{DEPLA2010253}, a $\SI{2}{\mu \mathrm{m}}$ thick layer of aluminum of ultra‑high purity was deposited on top of the wafers. The purity of metal targets in PVD chambers is a critical factor in ensuring high conductivity of the deposited films \cite{ashcroft1976solid}. Pure aluminum is suitable as an electrode metal due to its low specific resistance at cryogenic temperatures ($\SI{4.3e-10}{\Omega \mathrm{m}}$ at \SI{10}{K} \cite{Dietl2025}).

% lithography
The process of structuring the aluminum layer involved optical lithography steps succeeded by plasma etching \cite{Chapman1980GlowDP, Donnelly2013PlasmaEY}. The lithography process consisted of three stages: resist spinning, exposure to \SI{365}{nm} light, and resist development \cite{ZHANG20141}. A mask containing the chip design was repeatedly aligned and exposed across the wafer \cite{935480, 10.1117/12.240936}. 

Automated optical inspection \cite{2020AutomatedSF} examined every individual trap on the wafer for any defects. The process captured images of the entire wafer, chip by chip, and created a reference image by averaging the color values of each pixel within the red-green-blue (RGB) color vector space. The evaluation of individual chips occurred by calculating the distance between their pixel values and the averaged pixel values of the reference in RGB space. Any chip exceeding a defined threshold for this distance received a flag and did not progress to subsequent processing steps. The chosen threshold distance in RGB space achieved a balance between sensitivity to the smallest detectable defects and the risk of wrongly excluding chips without visible issues. This method enabled detection of defects with dimensions as small as several tens of micrometers. A mechanical dicing process \cite{Lei2012DieST} separated the wafer into the individual chip traps.

\subsection{Quantum processing unit and experimental setup}

\begin{figure*}
\centering
\includesvg[scale=0.29]{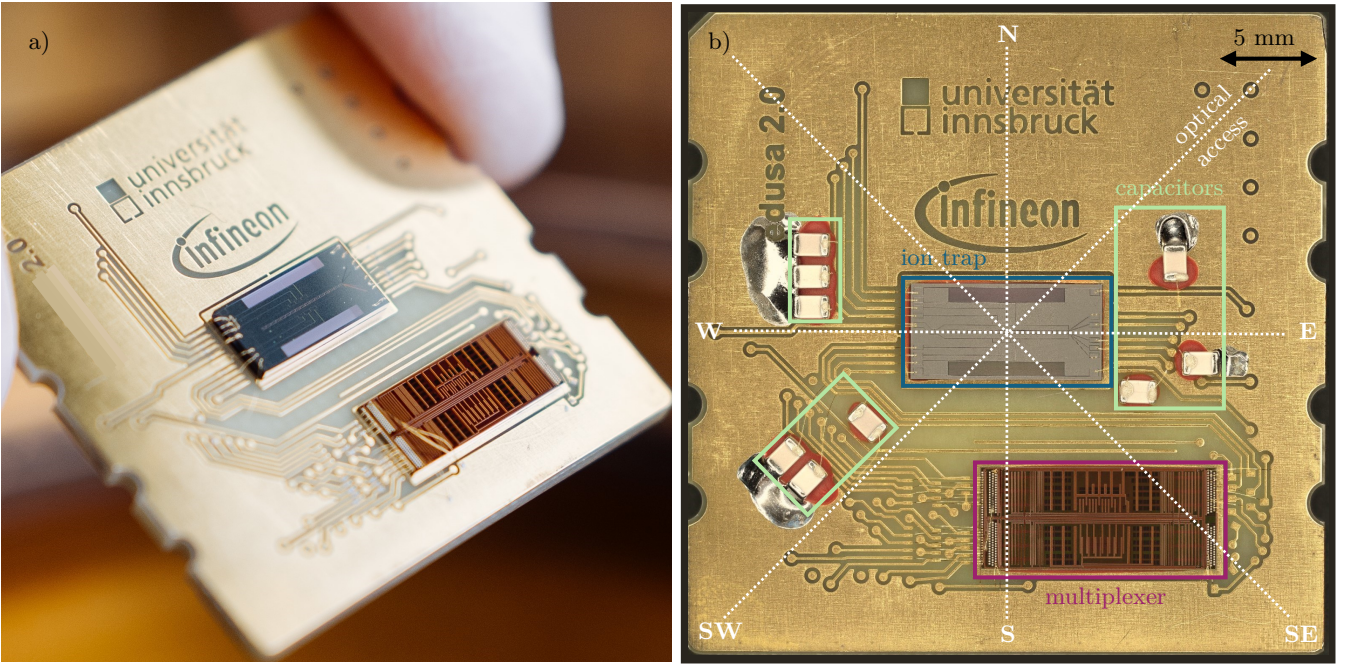}
\caption{\label{qpu} Quantum processing units featuring a surface ion trap at their centers and a multiplexer located in the south of the PCBs. Panel a) shows the first version of the QPU, without any additional capacitors mounted. Panel b) shows the second-generation design, which incorporates \SI{39}{nF} capacitors between the electrode lines and the PCB ground plane.
}
\end{figure*}

\begin{figure*}
\centering
\includesvg[scale=0.29]{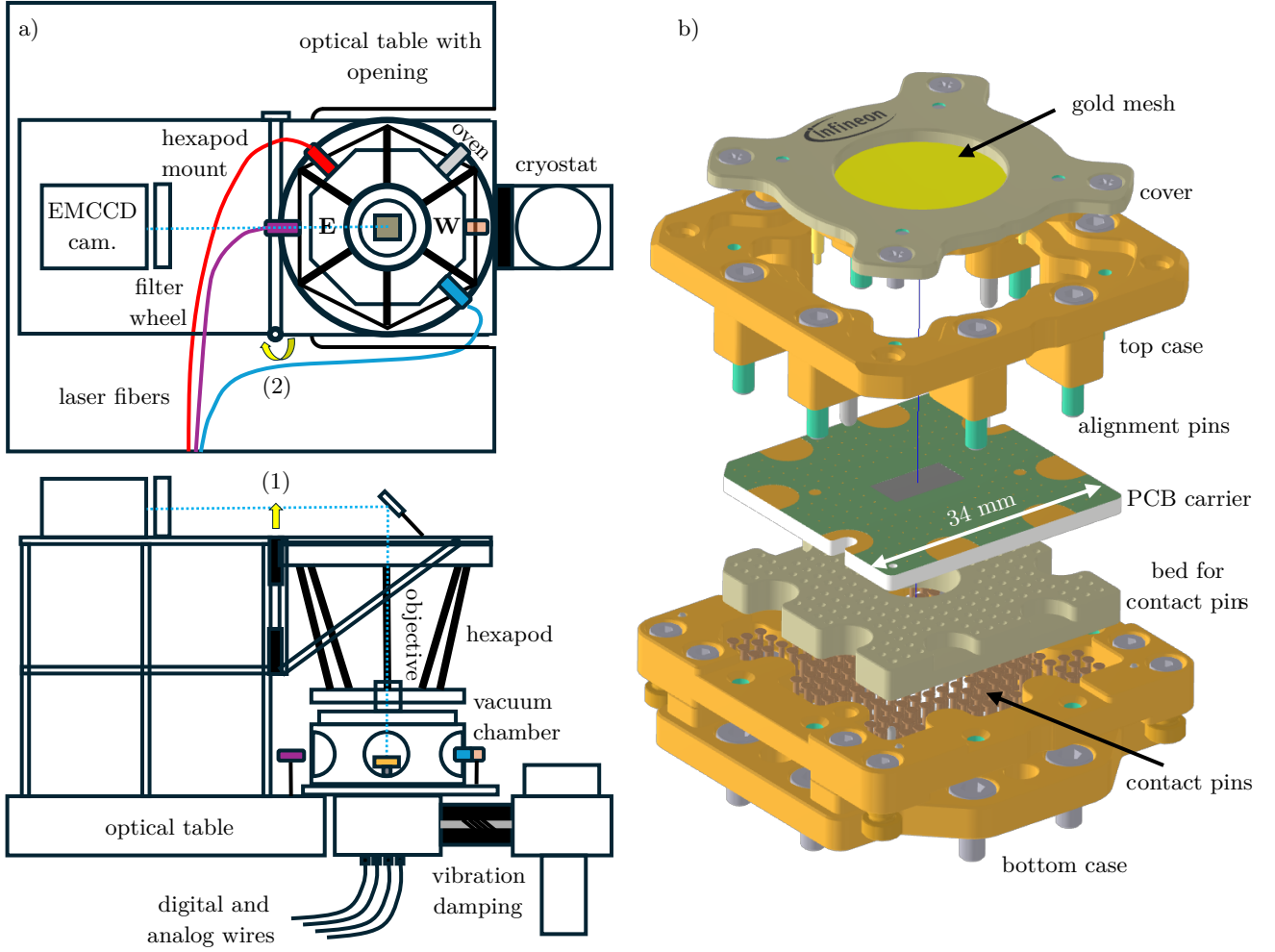}
\caption{Experimental setup and socket. a) Sketch of an optical table supporting a vacuum chamber, an EMCCD camera, a hexapod holding an optical objective. Yellow arrows indicate the movable range of the hexapod. b) Exploded view of the socket with a bottom case containing recessed beds for contact pins, which are inserted to provide electrical connections to a PCB carrier. The top case uses alignment pins for precise assembly, and a cover incorporates a gold mesh electrically connected via the PCB carrier.}
\label{fig:setup}
\end{figure*}

% qpu
The core of this work is the QPU consisting of the trap chip and the multiplexer, both assembled onto a printed circuit board (PCB), shown in Figure \ref{qpu}. The two chips were glued onto the square PCB with a base length of \SI{34}{mm} using an one-component epoxy adhesive, with the multiplexer positioned in the South (S) of the PCB and the ion trap placed in the center of the PCB. Gold wire bonds with a diameter of $\SI{50}{\mu \mathrm{m}}$ establish electrical connections between the chips and the lines on the PCB. 

% caps
The operation of analog switches involves charge injection, occurring during transitions between on and off states (Section \ref{sec: qpu}). The process is driven by the interaction between the electrode's capacitance and the switch's parasitic capacitance. When the switch is opened or closed, the stored charges redistribute, altering the electrode's potential. Capacitors with a capacitance of \SI{39}{nF} between the corresponding signal lines of the DC electrodes and the ground plane were added to minimize the potential change due to charge injection (Figure \ref{qpu}\textcolor{blue}{b}). As capacitance increases, the potential drop caused by charge injection decreases. With \SI{39}{nF}, the capacitance increases by three orders of magnitude compared to the scenario without additional capacitors. For a qubit and electrode count below approximately hundred, components with this level of capacitance remain compact enough to be integrated onto the PCB. The capacitors were arranged, such that optical access with lasers was guaranteed in the directions E-W, N-S, SW-NE and SE-NW. 

% laser access and laser delivery
With the \SI{39}{nF} capacitors in place, we proceed to laser access for ion control, directing the following lasers to the trap center: ionization lasers at wavelengths of \SI{375}{nm} and \SI{423}{nm}, a laser at \SI{397}{nm} for detection and Doppler cooling, repumping lasers at \SI{866}{nm} and \SI{854}{nm} and a qubit laser at \SI{729}{nm}. We delivered the necessary lasers to the ion through viewports in the cryostat chamber, each separated by \SI{45}{degrees}. The orientation of the QPU with respect to the chamber is indicated by cardinal directions (N, E, S, W) in Figures \ref{qpu}\textcolor{blue}{b} and \ref{fig:setup}\textcolor{blue}{a}. Specifically, the ionization and repumping lasers were placed along the trap's axial direction, coming from the east (E) direction. The \SI{397}{nm} laser for Doppler cooling and detection was inserted from the northwest (NW) direction, and the qubit laser from the southeast (SE) direction. 

% image description setup
The QPU is housed within a closed-cycle pulse tube cryostat, which provides a cryogenic environment with a base temperature of around \SI{10}{K} \footnote{Model: T-Type Optical}. The chamber is rigidly fixed to the table and connected to the cryostat via a flexible hose. A structure on top of the optical table supports an EMCCD camera \footnote{ANCOR iXon Ultra 888 CCD}. Fluorescence light at \SI{397}{nm} from the ion is guided to the camera via a custom objective \footnote{Photon Gear, NA=0.5, \SI{27}{mm} working distance} held by a hexapod \footnote{Physik Instrumente H-840.G2IHP}, a mirror and a filter wheel. 

% fast swapping
Building on the experimental setup described above, the development of ion QPUs is facilitated by rapid iteration cycles between design and characterization of surface ion-traps. Cryogenic systems operating below \SI{10}{K} eliminate the need for chamber bake-out procedures required by traditional room temperature setups, as gas molecules condense and freeze on the chamber's surface \cite{Ob_il_2019, Hahn2025ARE}. Thermal cycles can be completed in just a few hours. To facilitate quick QPU swaps, we've optimized the setup with a swiveling apparatus that moves the objective. As shown in Figure \ref{fig:setup}\textcolor{blue}{a}, the hexapod translates upwards (1) and then rotates to the side (2), enabling chamber access within minutes.

% socket
We have developed a socket, facilitating the packaging of chip traps (Figure \ref{fig:setup}\textcolor{blue}{b}). That housing can mount and electrically connect a QPU module on a square PCB with a side length of \SI{34}{mm} \footnote{Figure \ref{fig:setup}\textcolor{blue}{b} illustrates the second iteration of the socket, which features enhancements including alignment pins and an increased contact area between the PCB carrier and the socket to enhance thermal conductivity. However, it was the first version of the socket that was utilized in the experiments presented herein.}. The socket and the PCB carrier are thermally coupled, enabling the transfer of heat away from the trap chip. The QPU is placed in between two metal cases. The 172 electrical connections between the bottom case and the backside of the carrier PCB are implemented using contact pins \cite{Carter2000FuzzBI}, which are small, spring‑loaded connectors featuring a compliant, multi-filament surface \cite{fuzz_buttons}. The wires of that connectors consist of a beryllium copper core plated with a layer of gold, which are non-magnetic. Dedicated mounting holes support the cylindrical pins to enable reliable, solder‑free module exchange. The absence of soldering connections reduces the QPU exchange time. A gold mesh, consisting of $\SI{19}{\mu m}$ thick gold wires spaced $\SI{363}{\mu m}$ apart at each node, is suspended \SI{7.8}{mm} above the surface of the trap. This mesh is used to create a well-defined electric potential in the surrounding area \cite{Brandl2016CryogenicSF}, and is electrically connected to an external DAC.

% electrics
The organization and integration of the experiment’s electronic control components are essential for managing signal generation and delivery. The server to control the experiment and the Sinara hardware are integrated in a dedicated 19‑inch rack. Sinara is a modular electronics platform that interfaces with the experiment peripherals via the ARTIQ software \cite{Kasprowicz2020ARTIQAS, Przywzki2023SinaraAA}. The required RF trap drive and the RF signals for the double‑pass acousto‑optic modulators\footnote{Alpine Quantum Technologies, model Rowan} are generated by the Sinara Urukul, an RF source. The DC voltages used for axial ion confinement are generated by the Sinara Fastino module, a multi‑channel DAC. A first-order low-pass filter for each DC electrode with a cut-off frequency of \SI{20}{kHz} minimizes technical noise on the electrodes. These filters, comprising a $\SI{250}{\Omega}$ resistor and a \SI{33}{nF} capacitor, are mounted on a PCB placed inside the vacuum chamber at cryogenic operating temperatures, minimizing Johnson noise. The multiplexer operation necessitates independent DC voltage sources. Digital and analog signals to the physics chamber are delivered via four 25-pin D-SUB cables routed to the base of the system. The multiplexer is controlled by the digital signals from Sinara, which are transmitted using the LVDS \footnote{Low-Voltage Differential Signaling} protocol. Twisted pair wires, which consist of two insulated copper conductors wound together to reduce electromagnetic interference, are used within the cryostat to transmit the digital signals required for multiplexer control. In Reference \cite{Zahra2025}, the operational clock speed of the multiplexer was tested up to \SI{50}{MHz}. However, signal integrity constraints in our experimental setup led to communication failures with the ASIC at clock speeds exceeding \SI{2}{MHz}, and a conservative clock speed of \SI{500}{kHz} was therefore used to ensure robust operation throughout this study.

% However, in our experimental setup, signal degradation was observed at clock speeds exceeding \SI{2}{MHz}, leading to communication failure with the ASIC. A reduced clock speed of \SI{500}{kHz} ensured robust operation during this study.

\section{Trap and system characterization}
\label{sec:results}

This section reports on the experimental results regarding switching dynamics and heating rates obtained using the two QPU configurations illustrated in Figure \ref{qpu}, which were investigated in the research laboratory at Infineon Technologies, Villach. One QPU did not have the additional capacitors with \SI{39}{nF} soldered onto the PCB as shown in Figure \ref{qpu}\textcolor{blue}{a}. In this configuration, the effects of charge injection and voltage decay for open switches of the bare multiplexer were measured directly using a trapped ion. The voltage decay for electrodes with open switches, was studied.

For the second QPU, the \SI{39}{nF} capacitors between each DC electrode and ground have been added as shown in Figure \ref{qpu}\textcolor{blue}{b}. Consequently, the effect of charge injection onto the ion was minimized to a level, which was not measurable anymore. Heating rates for both, closed and open switch configurations were measured. 

\subsection{Metal temperature}

\begin{figure}
    \centering
    \includegraphics{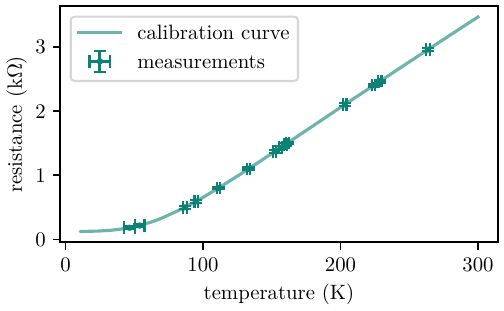}
    \caption{Temperature dependence of the thin film resistor.}
    \label{calib_curve}
\end{figure}

Characterizing trap performance as a function of electrode metal temperature provides valuable insights into the surface properties of the electrode metal and their underlying physics \cite{Boldin2017MeasuringAH}. The thin film resistors integrated into the chip allow to measure and control the temperature of the electrode metal. During the cool-down of the cryostat, the thin film resistor's resistance decreased from $\SI{3.2 +- 0.02}{\mathrm{k}\Omega}$ at \SI{300 +- 2}{K} to $\SI{42 +- 3}{\Omega}$ at \SI{10 +- 1}{K}, where the temperature was measured by a silicon diode mounted to the \SI{4}{K} shield of the cryostat. We assume that, due to a thermal link between the \SI{4}{K} shield and the trap socket, the temperature of the \SI{4}{K} shield is a good indicator of the trap's temperature. Figure \ref{calib_curve} shows the resistance of the thin film resistor as a function of the shield temperature. This temperature dependence was recorded while slowly cooling the cryostat, allowing sufficient time for the electrode metal temperature to equilibrate with the shield temperature. The resulting profile is consistent with the expected behaviour arising from established mechanisms governing metal resistivity \cite{Josell2009SizeDependentResistivity}. The data points correspond to the metal temperatures at which ions were trapped and QPU characterization was performed. Notably, activating the RF drive at a voltage amplitude of \SI{205}{V} resulted in a metal temperature increase from \SI{10 +- 1}{K} to \SI{47 +- 5}{K}. We suspect the bottleneck of thermal conductivity to be the layer of glue (one component epoxy) between the fused silica substrate and the metal of the PCB. 

A single thin film resistor on the chip is able to heat the metal temperature up to \SI{230}{K}. The resistance of the second thin film resistor was measured, to monitor the metal temperature. To reach temperatures above \SI{230}{K}, we had to heat with both thin film resistors simultaneously. In this case, the resistance was determined from the measured voltage and current. This way, the ion trap could be operated up to metal temperatures of almost \SI{300}{K}.

\subsection{Switching performance and floating electrodes}

% introduction
A promising concept to realize scalable trapped ion architectures is to operate the multiplexer in a sample-and-hold technique, described in Section \ref{subsec:WISE}. To maintain trapping conditions, the potential must remain stable even when the electrodes are switched to a floating state and their potential decays due to leakage currents. Thus, the switching behavior of the QPU was analyzed by selectively opening and closing the switches connected to different electrodes. We first analyze the effect of charge injection and voltage decay for floating electrodes, using a QPU without additional capacitors (Figure \ref{qpu}\textcolor{blue}{a}) on the PCB carrier. For these experiments, only the inner DC electrodes (1–5) located between the two RF rails were utilized for these experiments. Switches were initially closed, which means the electrodes were connected via the multiplexer to the DACs outside the cryostat. Once the switch was opened, the connected electrode was no longer connected to the DAC, but instead floated with respect to trap ground, with a capacitance of \SI{50}{pF} incorporated into the multiplexer.

\begin{figure}
    \centering
    \includegraphics{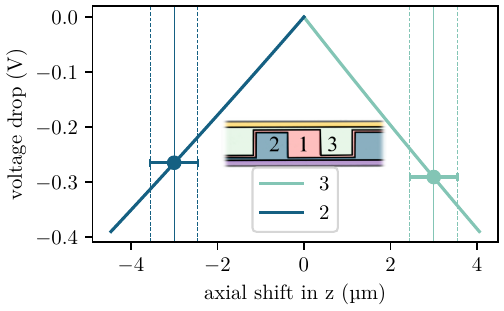}
    \caption{Voltage drop on electrodes 2 (blue) or 3 (green) vs the axial position of the ion. The two points indicate measured axial shifts of the ion due to charge injection, when the switches for electrode 2 or 3 were opened. The solid line shows the simulated behavior. Electrode inset is shown for orientation and is not to scale.}
    \label{fig:ion_jumps} % Voltages at $z=0$ are the applied voltages in the experiment.
\end{figure}

% charge injection
In the case of an ideal switch, the electrode potential would remain unchanged once a switch is opened. However, the charge injection effect (Section \ref{sec: qpu}) leads to a change in the electrode potential, which yields positional shifts of the ion. These shifts were observed whenever the switch to the respective electrode was opened or closed. In Reference \cite{Zahra2025} the multiplexer was characterized on its own and an estimation for the potential drop on a real electrode caused by charge injection was provided. Based on these results, a potential drop on the order of \SI{0.5}{V} would be expected for the electrode potential used in this trap. Switching off electrode 3 (2) led to a positional shift of the ion of $\SI{3 +- 0.5}{\mu m}$ ($\SI{-3 +- 0.5}{\mu m}$). Figure \ref{fig:ion_jumps} presents simulations of the axial ion position as a function of varying voltages for electrodes 2 or 3. The other electrode voltages (1, 4, 5) are held constant for the simulation. A comparison between the observed positional shifts and the voltage simulations reveals that the opening of switches results in a potential drop of \SI{0.29 +- 0.05}{V} (\SI{0.27 +- 0.05}{V}) for electrode 3 (2) with a nominal voltage of \SI{2.67}{V} (\SI{2.2}{V}).

\begin{figure}
    \centering
    \includegraphics{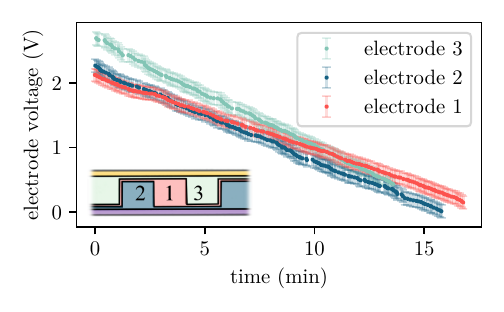}
    \caption{Estimated electrode voltages as a function of elapsed time after opening the switch of electrodes 1, 2, or 3. The fitted voltage decay rates are \SI{0.14739 +- 0.00041}{V/min}, \SI{0.16418 +- 0.00043}{V/min}, and \SI{0.12573 +- 0.00022}{V/min} for electrodes 1, 2, and 3, respectively.}
    \label{fig:electrode_decay}
\end{figure}

% voltage decay single
After the switch to a certain electrode was opened, the respective electrode was floating and its voltage decayed. This decay of potential for electrodes with open switches was measured by monitoring the ion's position as a function of time and comparing it to simulations, as illustrated in Figure \ref{fig:electrode_decay}. The voltage decay on each of the three inner electrodes was measured individually. The ion was positioned above electrode 1 to investigate the voltage decay of the adjacent electrodes 2 and 3. Since a decaying voltage on electrode 1 would displace the ion along the y-direction, parallel to the detection path, its position cannot be reliably measured with the EMCCD camera. The ion was therefore shuttled to a position above electrode 2, where the resulting axial displacement of decaying voltage on electrode 1 could be reliably detected. Although an exponential decay is expected from theory, the results show a linear decay with rates of \SI{0.14739 +- 0.00041}{V/min}, \SI{0.16418 +- 0.00043}{V/min}, and \SI{0.12573 +- 0.00022}{V/min} for electrodes 1, 2, and 3, respectively.

\begin{figure}
    \centering
    \includegraphics{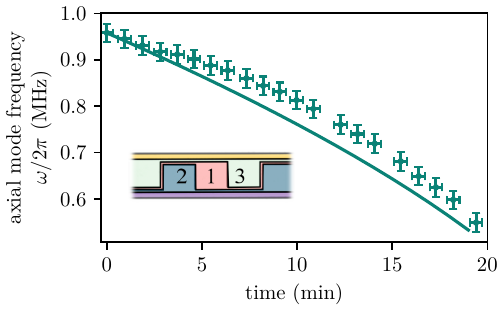}
    \caption{Axial mode frequency as a function of elapsed time after opening the switches of electrodes 2 and 3. Experimental data is shown as dotted points. The solid line indicates a simulated voltage decay rate of \SI{0.08 +- 0.02}{V/min}.}
    \label{voltage_decay}
\end{figure}

% voltage decay multi
The axial secular frequency of the ion was measured while electrodes 2 and 3 were left floating (ion trapped above electrode 1) to evaluate if the decay rates for individual switch openings are consistent with the case involving multiple open switches. The axial secular frequency was determined by applying an oscillating voltage to shim electrode 11 (Figure \ref{medusa_layout}\textcolor{blue}{b}) and varying the frequency around the expected 1$\times 2 \pi$ MHz resonance. When the excitation frequency matched the axial secular frequency, resonant axial oscillations of the ion were induced, allowing the axial secular frequency to be tracked as a function of time \cite{Home2011NormalMO, BautistaSalvador2018MultilayerIT}. These measured frequencies were again compared to simulations (Figure \ref{voltage_decay}), revealing a fitted linear voltage decay rate of \SI{0.08 +- 0.02}{V/min} for electrodes 2 and 3 being floating. For the sake of simplicity the same decay rate for both electrodes was assumed. The ion's lifetime was found to be around \SI{20}{minutes}, at which point it was lost. With the decay of \SI{0.08 +- 0.02}{V/min}, the potential at the time of ion loss have been, $\mathrm{DC}_1=\SI{-2.02 +- 0.02}{V}, \mathrm{DC}_2=\SI{0.68 +- 0.02}{V}$ and $\mathrm{DC}_3=\SI{2.15 +- 0.02}{V}$ corresponding to an axial frequency of approximately 0.5$\times 2 \pi$ MHz.

% summary
In summary, our measurements revealed voltage decay rates ranging from \SI{0.08}{V/min} to \SI{0.14}{V/min} for electrodes 1, 2, and 3. The origin of the observed variations in decay rates across the four different measurements remains unclear. One possible explanation could be the coupling of floating electrodes to their neighboring electrodes through parasitic capacitances and leakage currents. The capacitances between the DC electrodes are on the order of few hundreds of femtofarads (Table \ref{tab:caps_table}), whereas  each electrode is connected to ground through an integrated capacitance of \SI{15}{pF} provided by the multiplexer.

\begin{table}
    \centering
    \caption{Capacitances between next neighboring electrodes, calculated using finite-element analysis.}
    \begin{tabular}{|c|c|c|c|c|c|c|c|}
    \hline
        electrodes & GND-RF & RF-4 & 4-3 & 3-1 & 1-2 & 2-5 & 5-RF \\ \hline
        Capacity / pF & 1.1 & 0.22 & 0.20 & 0.32 & 0.32 & 0.20 & 0.22 \\ \hline
    \end{tabular}
    \label{tab:caps_table}
\end{table}

The voltage decay rates discussed above induce temporal variations in the trap frequency. These frequency fluctuations represent a critical source of error for high-fidelity qubit gates \cite{Ballance2017HighFidelityQL}, as the resulting detuning errors affect both laser-driven and microwave-driven gates. Low decay rates are therefore an essential requirement for architectures utilizing the sample-and-hold technique. Specifically, for electrodes with a capacitance to ground of \SI{50}{pF}, the observed decay rates of approximately $\SI{10}{\mu V/ms}$ correspond to a change in axial secular frequency of around $\dot{\omega}_{\mathrm{sec}}=\SI{0.4}{Hz/ms}$. Reference \cite{Ballance2017HighFidelityQL} identifies the individual sources of infidelity in two-qubit geometric phase gates \cite{Leibfried2003ExperimentalDO}. Among all these contributions, the gate detuning error $\epsilon_\delta$ is uniquely sensitive to the secular frequency of the ion on relevant time scales. The gate detuning error scales with the square of the absolute detuning frequency error $\dot{\omega}_{\mathrm{sec}} t$,

\begin{equation}
    \epsilon_\delta(t) \propto (\dot{\omega}_{\mathrm{sec}} t)^2\, T_G^2,
\end{equation}
where $T_G$ is the gate time. When sampling the voltages at frequencies above \SI{20}{Hz}, the change in secular frequency stays below \SI{20}{Hz} and the gate infidelity stays below $10^{-4}$ \cite{Hughes2025TrappedionTG} (for the estimation we assumed, $T_G=\SI{500}{\mu s}$).

% \begin{equation}
%     \epsilon_\delta(t) = \frac{1+2K\,(1+2 \bar{n})}{16K^2} \,(\dot{\omega}_{\mathrm{sec}} t)^2\, T_G^2,
% \end{equation}

\subsection{Charge Injection Suppression}

% problem
Charge injection displaced the ion by several micrometres, disrupting the spatial overlap with the laser beams required for cooling and qubit manipulation. Since precise laser–ion alignment is performed with closed switches, the ion displacement occurring upon opening the switches makes it impractical to perform optical operations on an ion confined by floating electrodes. However, the sample-and-hold operation inherently involves floating electrodes as part of the confining voltage sets and we aimed to characterize the QPU under both closed and open switch conditions.

% Therefore, it is essential to ensure low heating rates, when the switches are in the open state. % To reliably measure heating rates with open switches, it is necessary to suppress any ion displacement induced by charge injection. 

% solution
The addition of extra capacitors of \SI{39}{nF} per electrode to the PCB (Figure \ref{qpu}\textcolor{blue}{b}) led to an increase in shunt capacitance to ground by three orders of magnitude. As derived in Section \ref{subsec:multiplexer}, this reduces the potential drop caused by charge injection by a factor of thousand. After the addition of the capacitors, no positional shifts of the ion were observed, when opening or closing a single switch. Even when all switches were opened simultaneously, no shift in position of the ion occurred. We observed that ion trapping remained possible even after the switches had been open for approximately \SI{10}{hours}, whereas ions were lost for the case without additional capacitances after around \SI{20}{minutes}. From these maximum trapping times, we find the upper limit for the voltage decay for electrodes coupled to ground via \SI{39}{nF} to be \SI{2.5}{mV/min}. 

\subsection{Heating rates for closed,- and open switches}

% heating rates frequ
A fundamental limitation of trapped-ion quantum processors is motional heating, whereby electric-field noise from the trap electrodes limits the fidelity of qubit gates. In order to measure heating rates for the QPU with additional capacitors of \SI{39}{nF}, trapped ions were cooled close to the motional ground state using Doppler and sideband cooling. The average phonon number $\bar{n}$ was measured using sideband thermometry \cite{Leibfried2003} on the $4S_{1/2} \leftrightarrow 3D_{5/2}$ transition. The excitation probabilities of the red and blue sideband, $p_e^{\mathrm{red}}$ and $p_e^{\mathrm{blue}}$, were measured to determine $\bar{n}$ according,

\begin{equation}
    \bar{n} = \frac{p_e^{\mathrm{red}}/p_e^{\mathrm{blue}}}{1-(p_e^{\mathrm{red}}/p_e^{\mathrm{blue}})}.
    \label{equ:sideband_thermo}
\end{equation}
The excitation probabilities were measured at various delays to calculate the rate of excitation of the ion motion. The uncertainty arising from quantum projection noise \cite{Tommy_diss} was taken into consideration \footnote{The propagation of uncertainty in Eq. (\ref{equ:sideband_thermo}) was quantified using the relation, $\Delta \bar{n} = \sqrt{((\partial_{\bar{n}} / \partial p_e^{\mathrm{red}}) \Delta p_e^{\mathrm{red}})^2 + ((\partial_{\bar{n}} / \partial p_e^{\mathrm{blue}}) \Delta p_e^{\mathrm{blue}})^2} $, which was applied to determine the standard deviation in $\bar{n}$.}. Heating rates were measured for both, closed and open switches in the following way: the measurement started with closed switches to compensate for excess micromotion \cite{Berkeland1998MinimizationOI}. Next, the heating rate measurement was employed, using sideband thermometry after various delays. Then, the switches were opened and a second sideband thermometry scan with the same delays was performed. 

\begin{figure}
    \centering
    \includegraphics{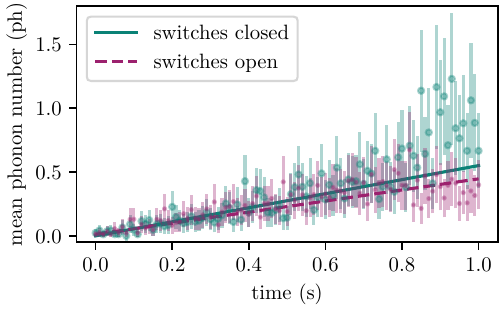}
    \caption{Mean phonon numbers for different delays and closed and open switches at 1.71$\times 2 \pi$ MHz axial frequency and \SI{47 +- 5}{K} metal temperature. A weighted linear fit was used to evaluate the heating rate from the motional occupation numbers. The heating rates are \SI{0.55 +- 0.21}{ph/s} for closed switches and \SI{0.43 +- 0.17}{ph/s} for open switches.}
    \label{lowest_heating_rates}
\end{figure}

\begin{figure}
    \centering    \includegraphics{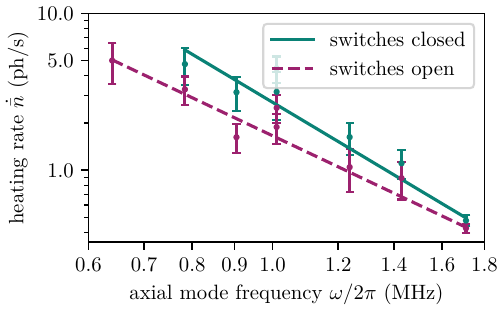}
    \caption{Heating rates as a function of axial frequency and a fit of the function $\dot{\bar{n}}(\omega) = \dot{\bar{n}}_{1}~\omega^{-\alpha}$ at \SI{47 +- 5}{K} metal temperature. The fitted exponent values are $\alpha = 3.16(28)$ for closed switches and $\alpha = 2.50(18)$ for open switches.}
    \label{frequ_dep_low}
\end{figure}

% lowest rates at 1.61 MHz
At a metal temperature of \SI{47 +- 5}{K} and an axial frequency of 1.71$\times 2 \pi$ MHz, the heating rate for closed switches was \SI{0.55 +- 0.21}{ph/s} and \SI{0.43 +- 0.17}{ph/s} in case of open switches, as shown in Figure \ref{lowest_heating_rates}. This measurement consisted of 100 points spanning from 0 to \SI{1}{s} waiting time and took roughly \SI{3}{hours}, over which the experimental system did not show substantial drifts. All other measurements consisted of 15 points, spanned between 0 and \SI{100}{ms} waiting time. Fitting the function $\dot{\bar{n}}(\omega) = \dot{\bar{n}}_1\omega^{-\alpha}$ to the experimental data yielded an exponent $\alpha$ of 3.16(28) for closed switches and 2.50(18) for open switches at a metal temperature of \SI{47 +- 5}{K}, as illustrated in Figure \ref{frequ_dep_low}.

At a metal temperature of \SI{47 +- 5}{K}, the heating rates for closed (open) switches were \SI{0.55 +- 0.21}{ph/s} (\SI{0.43 +- 0.17}{ph/s}) at 1.71$\times 2 \pi$ MHz axial frequency, as shown in Figure \ref{lowest_heating_rates}.

\begin{figure}
    \centering
    \includegraphics{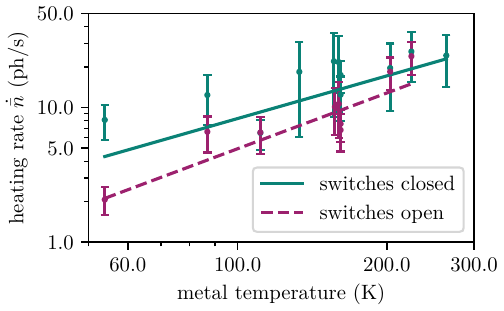}
    \caption{Heating rates as a function of metal temperature and a fit of the function $\dot{\bar{n}}(T) = \dot{\bar{n}}_0~(k_B T / \hbar \omega)^\beta$ at 0.75$\times 2 \pi$ MHz axial frequency. The fitted exponent values are $\beta=1.05(38)$ for closed switches and $\beta=1.37(29)$ for open switches.}
    \label{graph:temp_dep}
\end{figure}

% heating rates heating
To gain insight into the temperature dependence of the heating rate, the axial secular frequency was kept constant at 0.75$\times 2 \pi$ MHz and the metal temperature was varied. A representation of the data (Figure \ref{graph:temp_dep}) exhibits a linear correlation between the heating rate and the metal temperature, indicating a power-law dependence between these two parameters. This analysis is similar to Reference \cite{Bruzewicz2014MeasurementOI}, where heating rates were measured across metal temperatures from \SI{4}{K} to \SI{300}{K} and modeled using a power-law temperature scaling with a zero-temperature offset, $\dot{\bar{n}}(T) = \dot{\bar{n}}_0 (1 + (T/T_0)^{\beta})$. There, trap temperatures as low as \SI{4}{K} were achieved, our lowest measured metal temperature was \SI{47 +- 5}{K}. As a result, we lack heating rate data in the low-temperature regime and fit our data to the high-temperature approximation of the model, $\dot{\bar{n}}(T \gg T_0) \approx \dot{\bar{n}}_0~(\alpha T)^\beta$, with $\alpha = k_B/(\hbar \omega)$. The fit results show a scaling of $\beta=1.05 \pm 0.58$ for closed switches and $\beta=1.37 \pm 0.46$ for open switches, as presented in Figure \ref{graph:temp_dep}. 

% frequency dependence for different temperatures
To evaluate the temperature dependence of the $\alpha$ parameter in $\dot{\bar{n}}(T, \omega) = \dot{\bar{n}}_1(T, \omega) \omega^{\alpha(T)}$, heating rates were measured for varying frequencies also for \SI{95 +- 3}{K} and \SI{111 +- 2}{K} metal temperature. Corresponding graphs are provided in Appenidx \ref{app:full_data}, and all fit parameters are summarized in Table \ref{tab:exponent_fits}. Overall, the heating rates stayed well below \SI{10}{ph/s} at $1 \times 2 \pi$ MHz and metal temperatures up to \SI{110}{K}.
% For higher metal temperatures, we observe an increase in the fitted heating rate at $1 \times 2 \pi$ MHz, $\dot{\bar{n}}_1$, with a steeper increase for open switches ($\dot{\bar{n}}_{1}^{\mathrm{closed}}(\SI{111}{K}) / \dot{\bar{n}}_{1}^{\mathrm{closed}}(\SI{47}{K}) = 1.86$ and $\dot{\bar{n}}_{1}^{\mathrm{open}}(\SI{111}{K}) / \dot{\bar{n}}_{1}^{\mathrm{open}}(\SI{47}{K}) = 2.46$).

\begin{table}
    \centering
    \caption{Axial heating rates for closed and open switches at different metal temperatures and axial mode frequencies are fitted to $\dot{\bar{n}} = \dot{\bar{n}}_1 \omega^{\alpha}$.}
    \begin{tabular}{c|c|c|c}
        T (K) & switch configuration & $\dot{\bar{n}}_1(T)~(\mathrm{ph/s})$ & $\alpha(T)$ \\ \hline
        47(5) &  & 2.71(34) & 3.16(28) \\
        95(3) & closed & 4.37(34) & 2.74(36) \\
        111(2) &  & 5.04(17) & 2.59(17) \\ \hline
        47(5) &  & 1.66(14) & 2.50(18) \\
        95(3) & open & 2.93(21) & 1.53(30) \\
        111(2) &  & 4.08(20) & 1.70(22) \\
    \end{tabular}
    \label{tab:exponent_fits}
\end{table}

Summarizing the experimental results for the QPU configuration with added \SI{39}{nF} capacitors, charge‑injection effects were suppressed below measurable levels, and the voltage decay of floating electrodes was lower than \SI{2.5}{mV/min}. Heating rates as low as \SI{0.55 +- 0.21}{ph/s} were observed, with consistently lower values when the switches were open. The heating rate exhibited a power-law dependence on axial frequency, with scaling exponents of 3.16(28) for closed switches and 2.50(18) for open switches at a metal temperature of \SI{47 +- 5}{K}. For varying metal temperatures we found also a power-law scaling for the heating rates with exponents of 1.05(38) for closed and 1.37(29) for open switches.

\section{Fully integrated multiplexer}
\label{sec:outlook}

\begin{figure*}
    \centering
    \includesvg[scale=0.29]{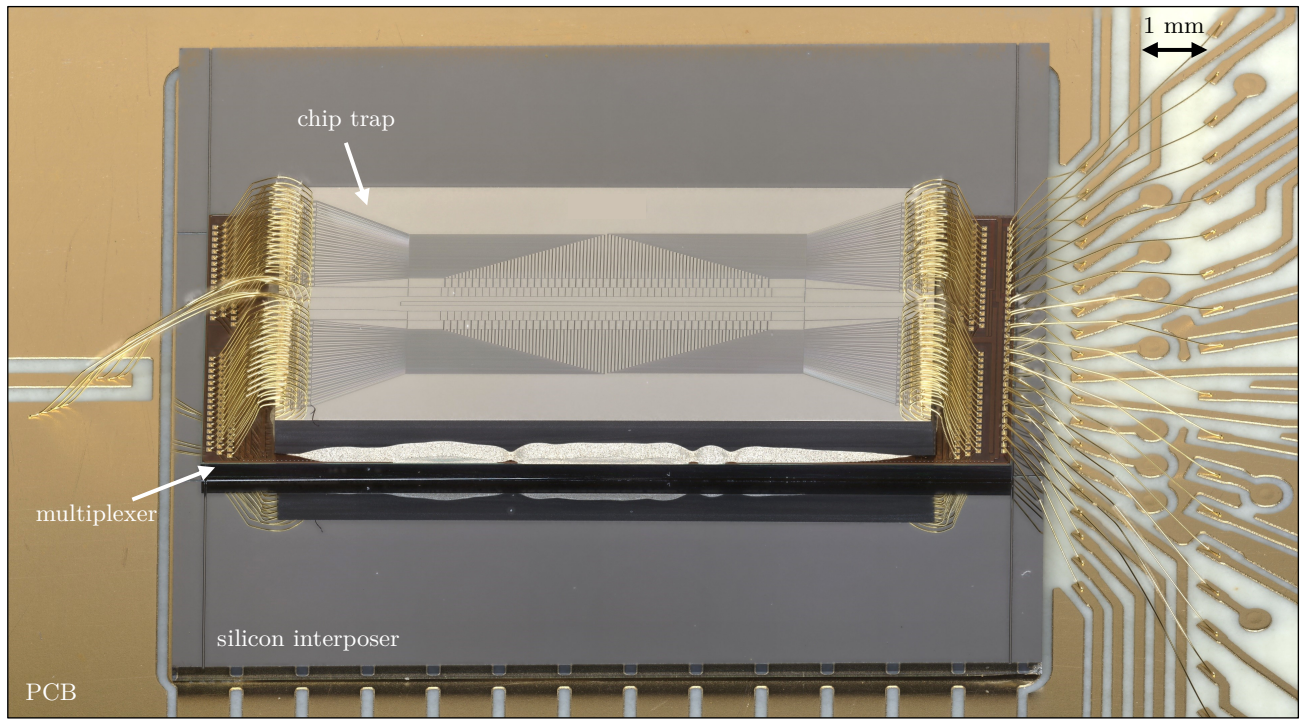}
    \caption{
    Stacked-chip QPU consisting of a multiplexer ASIC and the surface trap glued onto a silicon interposer. The multiplexer is connected to the PCB by wire bonds. The input voltages and digital control signals interface the PCB with the multiplexer on the right-hand side. On the left-hand side, the RF and trap ground connections link the PCB to the chip trap. The chip trap is glued on top of the multiplexer and interconnected to the multiplexer via 100 wire bonds per side, arranged in three rows.
    }
    \label{fig:amadeus_qpu_stack}
\end{figure*}

\begin{figure*}
    \centering
    \includesvg[scale=0.29]{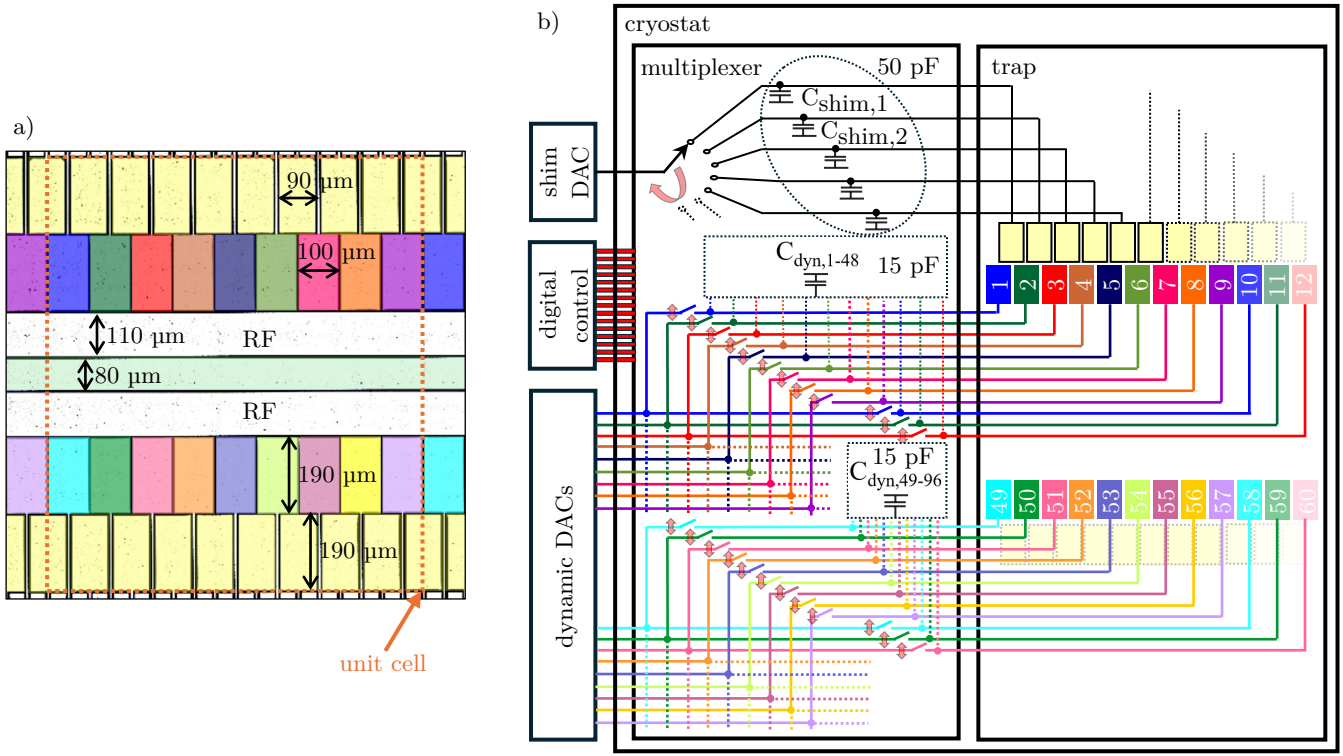}
    \caption{a) Colored microscope picture of the second traps electrode design. Shown is a complete unit cell of the design. Dynamical electrodes are shown in various colors and the shim electrodes in yellow. b) Wiring scheme of the stacked QPU. Dynamic DACs, a single shim DAC and the digital control of the multiplexer are placed outside the cryostat. Multiplexer and chip trap are inside the cryostat. Dynamic DACs (numbered) are co-wired with a periodicity of 9. Each dynamic electrode is connected to its DAC via a 1 x 1 switch and capacity to ground (\SI{15}{pF}) in the multiplexer. All shim electrodes (yellow) are connected to a single shim DAC via a 1 x N switching matrix and are connected to a capacity to ground (\SI{50}{pF}). 
    }
    \label{fig:amadeus_qpu_circuit}
\end{figure*}

% introduction
This section outlines a conceptual demonstrator of an enhanced QPU design, building upon the multiplexer architecture presented in this study. The multiplexer is capable of supplying up to 194 DC electrodes. The design and fabrication of a surface ion trap with 194 DC electrodes allowed for the full exploitation of the chip's potential, with the complete trap layout shown in Appendix \ref{app:amadeus}. The chip trap is shown as a component of a stacked QPU in Figure \ref{fig:amadeus_qpu_stack}. The electrode design and the electrical layout are presented in Figure \ref{fig:amadeus_qpu_circuit}. 

% trap geometry
In Figure \ref{fig:amadeus_qpu_circuit}\textcolor{blue}{a}, the unit cell of the trap layout is presented in more detail, highlighting the key dimensions and features of the electrode design. The RF rails with a width of $\SI{110}{\mu m}$ are separated by $\SI{90}{\mu m}$, yielding an ion-surface distance of $\SI{80}{\mu m}$, which is less than half compared to the initial design shown in Figure \ref{medusa_layout}. By reducing the proximity of the ions to the surface, smaller electrode footprints can be achieved while maintaining control voltages within the range of $\pm \SI{10}{V}$. Additionally, stronger electric field gradients achievable at smaller ion‑surface distances allow more precise control of the trapping potential, facilitating the splitting and merging of ion crystals. Nested between the RF rails, a $\SI{80}{\mu m}$ wide DC electrode is placed. All gaps separating the electrodes are $\SI{5}{\mu m}$ wide. The trap features 96 dynamic electrodes and 98 shim electrodes in a linear design. Next to the two RF rails, two rows of segmented dynamic electrodes are located. Dynamic electrodes have the dimensions $\SI{100}{\mu m} \times \SI{190}{\mu m}$. To the outside of the dynamic electrodes, the shim electrodes ($\SI{90}{\mu m} \times \SI{190}{\mu m}$) are placed. The trap design on its own has been successfully tested and characterized \cite{Werhounig2025}.

% wiring scheme
In addition to the trap design, the wiring scheme of the stacked QPU is an important aspect of the overall architecture. The wiring scheme is shown in Figure \ref{fig:amadeus_qpu_circuit}\textcolor{blue}{b}. Eighteen dynamic DACs are connected to two rows of dynamic electrodes via analog switches in the multiplexer. For both rows, there are 48 dynamic electrodes, which are co-wired with a periodicity of nine, such that every 9$^\mathrm{th}$ electrode is connected to the same DAC. The dynamic electrodes enable ion sorting operations at different trapping sites simultaneously. Nine electrode pairs are sufficient for shuttling and merging/splitting of ion strings \cite{Optimization} and ion rotations \cite{Mourik2020CoherentRO}. We added capacitances into the multiplexer for the dynamic electrodes with \SI{15}{pF}. 

% wiring architecture + shim electrodes
Local electric fields, which might vary from one unit cell to another are neutralized by compensation electrodes. The 98 compensation electrodes (yellow) are located on the outside of the dynamic electrodes. The voltage for all compensation electrodes is supplied by a single DAC in the sample-and-hold manner with \SI{50}{pF} to ground.

% stacked QPU
The implementation of this wiring design is constrained by the fabrication limits of PCB technology. Specifically the minimal feature size, results in bond pads for wire bonding that are relatively large. Consequently, attempting to establish almost 200 connections between the ASIC and the trap on the PCB becomes impractical due to the significant space requirements for the bond pads. As a result, the two chips are stacked on top of each other and are directly connected with wire bonds, as shown in Figure \ref{fig:amadeus_qpu_stack}. The PCB is connected to the multiplexer on the right-hand side via 15 digital,- and 22 DAC signals. A total of 194 wire bond connections were established between the multiplexer and the trap to supply the 194 DC electrodes. To accommodate the high density of wire bonds (ca. 50 wires per $\SI{0.8}{mm^2}$) required for the trap, with 97 wire bond connections on both sides, the bond pads were arranged in three columns on the trap and the multiplexer. The pitch between two bond wires is $\SI{115}{\mu m}$ and a semi-automated wire bonding machine was employed. Ground and RF connections were made directly from the PCB to the trap.

% Heating to 70 K
The prototype of the multiplexer is susceptible to switching transients that distort time-varying voltages when operated at temperatures below \SI{70}{K}, as reported in Reference \cite{Zahra2025}. The reason for this switching transients at cryogenic temperatures is the freeze-out of charge carriers in the transistors. Since no ion shuttling with the multiplexer was conducted by now, that characteristic did not play a role so far. To address this issue, we employ a heating mechanism that raises the temperature of the ASIC from \SI{10}{K} to \SI{90}{K} by dissipating heat through surface mount (SMD) resistors on the silicon interposer. This interposer is positioned directly beneath the multiplexer. Measurements obtained using diodes within the multiplexer reveal that a power dissipation of approximately \SI{400}{mW} at the resistor is sufficient to bring the ASIC within its working temperature range above \SI{70}{K}. The temperature of the electrode metal was found to closely track the temperature of the ASIC.

% glueing
In total, the QPU consists of three stacked layers mounted on a PCB (Figure \ref{fig:amadeus_qpu_stack}): a silicon interposer (silver), a multiplexer (brown), and a surface ion trap on top. The assembly of the QPU stack demands particular attention, as the employed adhesives must be compatible with both ultra-high vacuum conditions and cryogenic temperatures. To identify suitable adhesives for each interface, a comprehensive study was conducted, involving temperature cycles from \SI{300}{K} to \SI{70}{K} using liquid nitrogen, as well as to \SI{10}{K}. Shear tests were performed at each interface to determine the adhesion between the respective chips. A summary of the adhesive types used, and the corresponding adhesion forces is presented in Table \ref{table:interface_values}. As a result, we chose a silver die attach adhesive as it showed acceptable shear force densities for all three interfaces of minimum $\SI{3.28(2)}{N/\mathrm{mm}^2}$ after being cooled to \SI{10}{K}.

\begin{figure}
    \centering
    \includegraphics{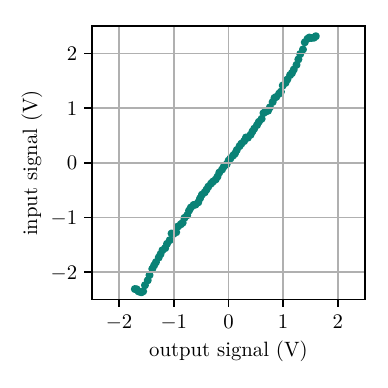}
    \caption{Input‑output characteristic for voltage signals going through the multiplexer at \SI{90 +- 3}{K} and an RF trap drive of $\SI{110}{V}$.}
    \label{graph:crymson_in_out}
\end{figure}

% RF-switch coupling
By stacking the ion trap on top of the multiplexer, the sensitive electronics of the multiplexer are now in closer proximity to the RF trap drive, raising concerns about potential interference from the oscillating voltages on the switches. To investigate this issue, a sinusoidal signal with an amplitude of \SI{2.5}{V} and a frequency of \SI{10}{kHz} was intentionally chosen and injected into one of the multiplexer's DC channels. The signal was then routed out of the cryostat via a dedicated debugging channel provided by the multiplexer. The multiplexer's temperature is affected by the RF trap drive and was additionally heated to the operational range above \SI{70}{K} by dissipating heat through the SMD resistor. At a multiplexer temperature of approx. \SI{90}{K} (measured with a diode integrated into the multiplexer), the signal integrity remained unaffected for RF amplitudes of $\SI{110}{V}$, which are needed for trapping ions. A decrease in signal strength to ca. $\SI{95}{\%}$ was present and is illustrated in the linearity plot in Figure \ref{graph:crymson_in_out}. This input-output characteristic was independent of the RF trap drive voltage ranging from \SI{0}{V} to \SI{110}{V}.
% The lower amplitude was selected to highlight the nonlinearity, which becomes less visible at higher voltages. Furthermore, the frequency of \SI{10}{kHz} was chosen to ensure that the wiring capacitances, which start to filter out the signal at higher frequencies, would not significantly affect the measurement.

\section{Conclusion}
\label{sec:conclusion}

% introduction
Scaling trapped‑ion systems is hindered by the need to route many individual DC voltages into a space‑constrained cryostat. This can be alleviated by integrating electronic switches into the trap module to enable time‑division multiplexing, where one line sequentially biases multiple electrodes. A sample‑and‑hold scheme maintains each electrode’s potential via a local capacitance, allowing several electrodes to be driven by a single DAC and substantially reducing wiring requirements.

% summary of this study
This study explores the integration of surface ion traps with electronic multiplexers, both operating at cryogenic temperatures, and demonstrates the feasibility of this approach. A key focus lies on understanding how electronic components impact the performance of the ion traps, with particular attention to their influence on ion heating rates. The research examines the principles of the sample-and-hold technique. Additionally, the study aims to establish robust and reliable methods for packaging these components onto PCBs and performing wire bonding, leveraging the capabilities of industrial manufacturing facilities.

% measurement results
Heating rates as low as \SI{0.55 +- 0.21}{ph/s} were measured, both for switches being closed and open. Hence, the multiplexer introduces negligible noise to the DC electrodes with a voltage noise amplitude of approximately $54~\mathrm{pV}/ \sqrt{\mathrm{Hz}}$. The heating rate, and thus the inferred electrode noise, was consistently lower when the switches were open. The dependence of the heating rate on the axial frequency, $\dot{\bar{n}}(\omega) = \dot{\bar{n}}_{1}~\omega^{-\alpha}$, scales with $\alpha = 3.16(28)$ for closed,- and $\alpha = 2.50(18)$ for open switches at a cryostat temperature of \SI{10 +- 1}{K}. While opening a switch, the respective potential at the electrode drops by around \SI{10}{\%} of its nominal potential. That potential drop leads to positional ion jumps in the order of µm. The voltage decay rate for electrodes with open switches was measured using ions and yielded around \SI{100}{mV/min} for \SI{50}{pF} capacity between electrode and ground and less than \SI{2.5}{mV/min} for \SI{39}{nF}. 

% assembly limitations
The assembly techniques outlined in this work involve attaching chips to a PCB using adhesives and establishing wire-bonded connections between the chips. While functional, this approach has significant scalability limitations. Future iterations of the design will address these challenges by employing through-substrate vias and bump bonding techniques to directly connect the electronics chip to the ion trap. Alternatively, the electronics could be monolithically integrated into the trap chips themselves \cite{Stuart2019, Slusher2005}.

% multiplexer limitations
The current multiplexer design has two significant limitations. Firstly, at temperatures below \SI{70}{K}, the multiplexer's output voltages exhibit transients, which can compromise the voltage sequences for ion crystal reconfigurations. In this work, we assembled the QPU in a way, such that the multiplexers can be heated up above \SI{70}{K}. Another strategy of coping with the transients would be a calibration of the switches' transfer function and modifying the input signal accordingly. The second substantial limitation of the current multiplexer design is the charge injection that occurs during switching. Within this work, the charge injection was suppressed by adding \SI{39}{nF} capacitors to each DC electrode. However, when performing fast shuttling or split/merge operations on ion crystals, high capacitances and serial resistances jointly constrain the achievable voltage ramps. The capacitances require high current amplitudes, while the resistances limit the maximum current, which slows the voltage transitions and hinders the realization of the necessary voltage profiles over time. Resolving the issue of charge injection will be critical for enabling reliable and efficient operation of the multiplexer in future applications. 

% multiplexer's decay rates, gates, ...
In addition to the limitations of the multiplexer previously discussed, our measurements revealed sufficiently low voltage decay rates for floating electrodes during qubit gates. The measured decay rates correspond to a secular-frequency drift of about \SI{0.4}{Hz/ms}, and since detuning errors scale with the square of this drift, sufficiently frequent voltage refresh cycles are required. Refreshing the voltages at rates above roughly \SI{20}{Hz} is compatible with a gate infidelity below $10^{-4}$.

\bibliography{literature}
\bibliographystyle{unsrt}

\appendix
% \onecolumngrid

\section{Full data set of heating rates}
\label{app:full_data}

\begin{figure*}
    \centering    
    \includegraphics{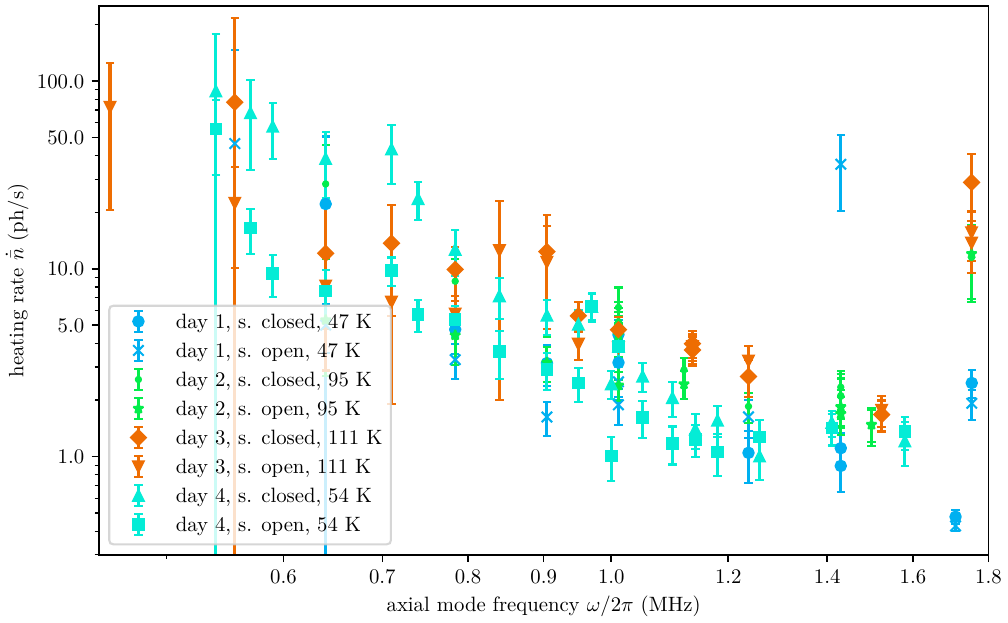} 
    \caption{All measured heating rates for different metal temperatures and axial secular frequencies.} 
    \label{graph:all_data}
\end{figure*}

\begin{figure}
    \centering    
    \includegraphics{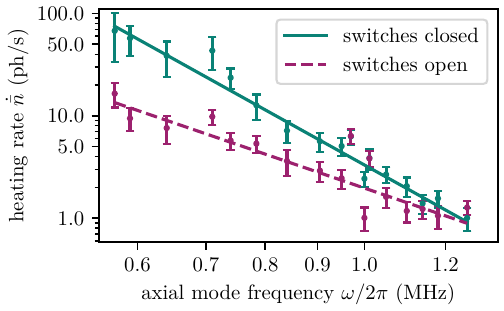} 
    \caption{Heating rates at different axial frequencies at \SI{54(5)}{K} metal temperature. A fit of the function $\dot{\bar{n}}(\omega) = \dot{\bar{n}}_{1}~\omega^{-\alpha}$, yields $\alpha=5.55(32)~\mathrm{and}~n_1 = 3.31(20) ~\mathrm{ph/s}$ for closed switches and $\alpha=3.42(42)~ \mathrm{and}~ n_1 = 1.97(21)~ \mathrm{ph/s}$ for open switches. This data was taken on day 4, where we observed a significant deviation in $\alpha$, compared to the other days of measurement. The reason for that additional noise stays unclear. However, this day, we were not able to reach the \SI{47}{K} base metal temperature again. Instead, the lowest metal temperature was \SI{54(5)}{K}.} 
    \label{frequ_dep_718}
\end{figure}

\begin{figure}
    \centering
    \includegraphics{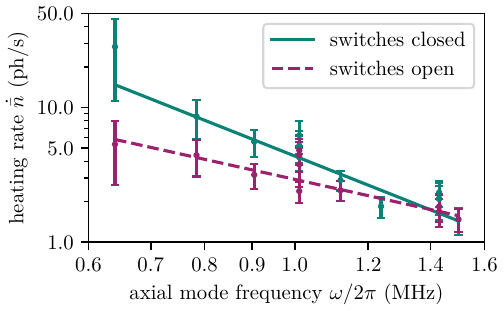}
    \caption{Heating rates at different axial frequencies at \SI{95(3)}{K} metal temperature. A fit of the function $\dot{\bar{n}}(\omega) = \dot{\bar{n}}_{1}~\omega^{-\alpha}$, yields $\alpha=2.74(36)~\mathrm{and}~n_1 = 4.67(34) ~\mathrm{ph/s}$ for closed switches and $\alpha=1.53(30)~ \mathrm{and}~ n_1 = 2.93(21)~ \mathrm{ph/s}$ for open switches (day 2).}  
    \label{frequ_dep_middle}
\end{figure}

\begin{figure}
    \centering
    \includegraphics{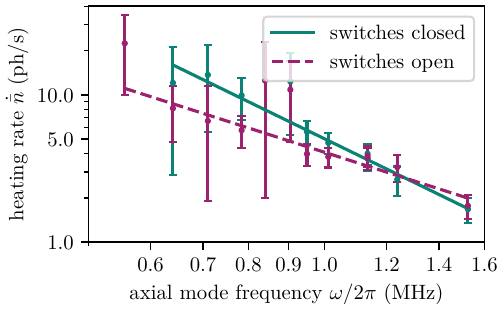}
    \caption{Heating rates at different axial frequencies at \SI{111(2)}{K} metal temperature. A fit of the function $\dot{\bar{n}}(\omega) = \dot{\bar{n}}_{1}~\omega^{-\alpha}$, yields $\alpha=2.59(17)~\mathrm{and}~n_1 = 5.04(17)~\mathrm{ph/s}$ for closed switches and $\alpha=1.70(22)~ \mathrm{and}~ n_1 = 4.08(20)~ \mathrm{ph/s}$ for open switches (day 3).}
    \label{frequ_dep_high}
\end{figure}

In Figure \ref{graph:all_data}, we provide the complete dataset of heating rates recorded across the four-day period, during which the frequency dependence was measured at varying metal temperatures. In the graphs presented in the text above, we removed outliers with significantly high standard deviations, promoting readability. Furthermore, we observed a day dependent noise for frequencies above $1.6\times2 \pi$ MHz, which are also not shown in the main text. Measurements of day 4 are shown in Figure \ref{frequ_dep_718}, but were excluded from table \ref{tab:exponent_fits}, since there was a substantial deviation in $\alpha$ in $\dot{\bar{n}}(\omega) = \dot{\bar{n}}_{1}~\omega^{-\alpha}$, compared to the other days of measurement. Figures \ref{frequ_dep_middle} and \ref{frequ_dep_high} show the heating rate at varying axial frequencies for \SI{95(3)}{K} and \SI{111(2)}{K}, respectively.

\section{Ion shuttling simulation}
\label{app:shim}

\begin{figure*}
    \centering
    \includegraphics[width=1\textwidth]{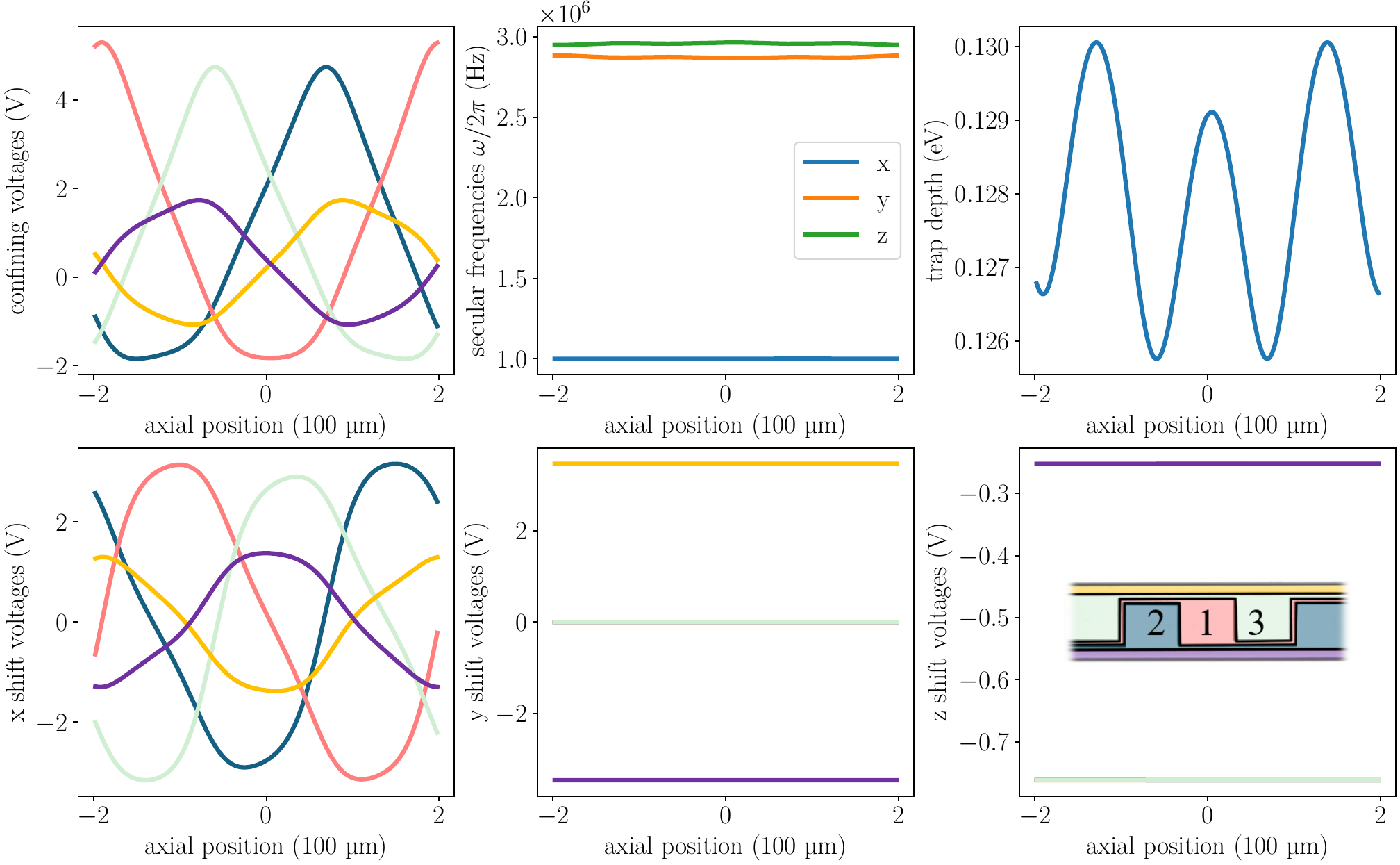}
    \caption{Ion shuttling simulations for trap type 1 ($\SI{170}{\mu m}$ ion-surface distance) over one periodicity of the periodic design of the electrodes. The first row shows the voltage set for confining ions, secular frequencies and trap depth. The color code of the voltage plots corresponds to the electrode colors shown in one sub graph. The sub graphs in the second row show DC voltages for compensating an electric stray field of \SI{1}{V/mm}. In the sub graph of the y shift, voltages of electrode 1, 2 and 3 are overlapping. In the sub graph of the z shift, voltages of electrode 1, 2 and 3 are overlapping at $\approx -0.8 V$ and the other two voltages are overlapping at $\approx -0.2 V$.}
    \label{fig:medusa_shimsets}
\end{figure*}

Figure \ref{fig:medusa_shimsets} shows the voltages, trap depth and secular frequencies for axial ion transport with the trap showed in Figure \ref{medusa_layout} with an ion-surface distance of $\SI{170}{\mu \mathrm{m}}$. Only the (dynamic) DC electrodes between the RF rails are used here. The supgraphs in the upper row (from left to right) show the confining voltages, secular frequencies and trap depth vs the axial ion position. Likewise, the bottom row shows the shim voltage sets for x, y and z electric stray field compensation.

\section{Trap layout for stacked QPU}
\label{app:amadeus}

\begin{figure*}
    \centering
    \includesvg[scale=0.15]{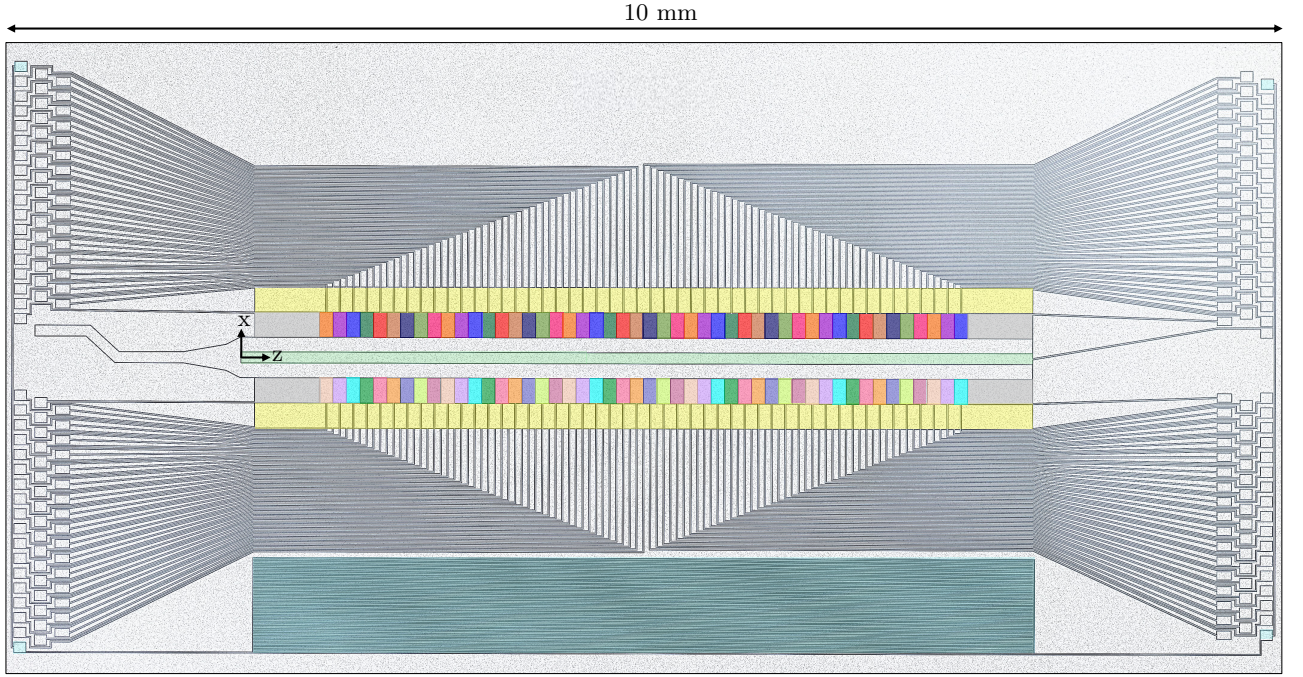}
    \caption{Microscope picture of the surface ion trap type 2 used for the stacked QPU, with chip dimensions of \SI{10}{mm} $\times$ \SI{5}{mm}. Electrodes are arranged in seven rows in a linear fashion. The elongated electrode in the z symmetry axis is a $\SI{80}{\mu m}$ wide DC electrode. Next to this centered DC electrode, the two RF rails are located with a width of $\SI{110}{\mu m}$. Outside the RF rails, the segmented dynamic DC electrodes with dimensions $\SI{100}{\mu m} \times \SI{190}{\mu m}$ are located. Every 9$^{\mathrm{th}}$ dynamical electrode in both rows is connected to the same DAC via the multiplexers' wiring. This periodicity is represented by the colors of the electrodes. Next to the dynamical electrodes, the compensation electrodes are located with dimensions, $\SI{90}{\mu m} \times \SI{190}{\mu m}$. The different levels of transparency in the yellow color indicate different voltages, provided by a single DAC in the sample-and-hold manner. All gaps between the features are $\SI{5}{\mu m}$ wide. The bond pads for connecting the electrodes to the multiplexer via wire bonds are arranged in three columns on the left and right sides of the chip. The blue rectangle at the bottom of the chip is a thin film resistor, which is connected to four bond pads.}
    \label{fig:amadeus_layout}
\end{figure*}

Figure \ref{fig:amadeus_layout} shows an optical microscope picture of the surface ion trap with $\SI{80}{\mu \mathrm{m}}$. 

\section{Adhesive study for QPU assembly}

Table \ref{table:interface_values} shows how strongly several adhesives bond at different layers within a multi‑part QPU stack after repeated cryogenic‑temperature cycling.

\label{app:glues}
\begin{table}
\centering
\caption{
Adhesive strength for various glues at three distinct interfaces within a stacked assembly as shown in Figure \ref{fig:amadeus_qpu_stack}. The interfaces examined are: (1) the gold (Au) surface of the printed circuit board (PCB) and the native oxide (SiOx) on the backside of the silicon interposer, (2) the silicon top surface of the interposer (SiOx) and the native oxide backside of the multiplexer (SiOx), and (3) the top of the multiplexer (Imide) and the native oxide on the back of the ion trap (SiOx). The glue column refers to the following adhesives: 1 - hybrid chemistry-based, non-conductive die-attach adhesive, 2 - silver die attach adhesive, 3 - one component epoxy adhesive, and 4 - two-component, solvent-thinned, epoxy-phenolic strain gage adhesive. After assembly and 10 cooling dips in liquid nitrogen, the interfaces were subjected to shear testing until die detachment, with the shear force monitored throughout. The maximum shear forces achieved after cooldown to \SI{10}{K} in the cryostat are listed in column 4. A star symbol (*) indicates that the maximum shear force reached its upper limit of \SI{1}{kN}, at which point the measurement was terminated. Uncertainties for all values are within \SI{5}{\%}.
} % Notes: samples 04 and 08 have been to cryo; 01 was definitely not, and it is unclear for 03; glues: TS333/TS336/TS336.
\begin{tabular}{|c|c|c|c|}
\hline
Interface & Glue & $\mathcal{F}_{\SI{70}{K}}$ & $\mathcal{F}_{\SI{10}{K}}$ \\
 &  & $\mathrm{N}/\mathrm{mm}^2$ & $\mathrm{N}/\mathrm{mm}^2$ \\
\hline
Imide - SiOx  & 1 & 7.32 & \\
Imide - SiOx  & 2 & 6.48 & \\

Imide - SiOx  & 2 & & 5.49* \\
Imide - SiOx  & 2 & & 5.18 \\
Imide - SiOx  & 2 & & 4.77 \\
Imide - SiOx  & 2 & & 5.49* \\

\hline

SiOx - SiOx  & 3 & 13.80 & \\
SiOx - SiOx  & 3 & 13.80 & \\
SiOx - SiOx  & 3 & 13.80 & \\
SiOx - SiOx & 2 & 5.55 & \\
SiOx - SiOx  & 2 & 15.09 & \\
SiOx - SiOx & 2 & 6.12 & \\
SiOx - SiOx  & 1 & 13.80 & \\
SiOx - SiOx  & 2 & 6.85 & \\
SiOx - SiOx & 2 & 3.26 & \\
SiOx - SiOx  & 1 & 15.33* & \\
SiOx - SiOx & 2 & 4.77 & \\
SiOx - SiOx & 1 & 15.33* & \\
SiOx - SiOx  & 1 & 15.33* & \\

SiOx - SiOx  & 2 &  & 3.28 \\
SiOx - SiOx  & 2 &  & 5.94 \\
SiOx - SiOx  & 2 &  & 5.44 \\
SiOx - SiOx  & 2 &  & 5.87 \\

\hline

Au - SiOx  & 3 & 5.48* & \\
Au - SiOx & 3 & 5.48* & \\
Au - SiOx & 4 & 5.48* & \\
Au - SiOx & 4 & 5.48* & \\
Au - SiOx & 2 & 1.61 & \\
Au - SiOx & 1 & 5.48* & \\
Au - SiOx & 2 & 3.85 & \\
Au - SiOx & 2 & 1.78 & \\
Au - SiOx & 1 & 5.48* & \\
Au - SiOx & 2 & 3.45 & \\
Au - SiOx  & 1 & 5.48* & \\
Au - SiOx & 1 & 5.48* & \\

Au - SiOx & 2 &  & 5.13 \\
Au - SiOx & 2 &  & 9.09 \\
Au - SiOx & 2 &  & 20* \\
Au - SiOx & 2 &  & 8.21 \\

\hline
\end{tabular}
\label{table:interface_values}
\end{table}

\end{document}